\RequirePackage[2018-12-01]{latexrelease}
\documentclass{gGAF2e}

\usepackage{xcolor}
\usepackage{comment}

\usepackage{bm}

\newcommand{\ds}{\displaystyle}

\renewcommand{\Delta}{\triangle}

\newcommand{\pdr}{\frac{\partial}{\partial r}}

\newcommand{\pdt}{\frac{\partial}{\partial t}}
\newcommand{\pdth}{\frac{\partial}{\partial \theta}}

\newcommand{\tck}{\textcolor{black}}

\begin{document}

\jvol{00} \jnum{00} \jyear{2026} 

\title{ Nonlinear interaction between dynamo-generated magnetic fields, mean flows and internal gravity waves in stellar stably-stratified layers: From 3D to 1D}

\markboth{Daniel \textit{et al.}}{ Wave-induced mean flow and dynamo }

\author{F.Daniel$^{a,b}$ $^{\ast}$ \thanks{ $^\ast$ Corresponding author. Email: florentin.daniel@northwestern.edu},   L.Petitdemange$^{c}$, C.Pinçon$^{d}$, K.Belkacem$^{c}$, B.Longo$^{e}$ and C.Gissinger$^{f}$,  \\
  \vspace{6pt}
  $^{a}$Laboratoire de Physique de l'Ecole normale superieure, ENS, Universite PSL, CNRS, Sorbonne Universite, Universite de Paris, Paris, France\\
  $^{b}$ Center for Interdisciplinary Exploration and Research in Astrophysics (CIERA), Northwestern University, Evanston, IL, USA \\
  $^{c}$ Laboratoire d’Instrumentation et de Recherche en Astrophysique (LIRA), Observatoire de Paris, Université PSL, Sorbonne Université, Université Paris Cité, CY Cergy Paris Université, CNRS, 75014 Paris, France \\
$^{d}$ Université Paris-Saclay, CNRS, Institut d’astrophysique spatiale, 91405, Orsay, France\\
$^{e}$ Departament d'Astronomia i Astrofisica, Universitat de Valencia, Dr. Moliner 50, 46100 Burjassot, València, Spain \\
$^{f}$ Laboratoire de Physique de l'Ecole normale superieure, ENS, Universite PSL, CNRS, Sorbonne Universite, Universite de Paris, Institut Universitaire de France (IUF), Paris, France \\
\vspace{6pt}\received{v4.4 released October 2012}}
\maketitle
\begin{abstract}

\tck{The large observational dataset currently available has revealed our ignorance regarding angular momentum redistribution during the stars' life.} Angular momentum (AM) transport in stars results from a variety of physical mechanisms in which \tck{internal gravity waves (IGW)} and magnetic fields are expected to play a role. 
\tck{While magnetic fields have been well constrained at the surface of several massive and intermediate-mass stars, their origin and properties in deep stellar radiative interiors are still debated, despite recent promising detections in the core of some red giant stars.}
Therefore, the modelling of AM transport in stellar radiative layers only relies on theoretical and numerical estimates of magnetic fields.
\tck{Recent 3D numerical simulations show that a dynamo could occur in such deep radiative regions.} A realistic setup for understanding AM transport in such layers thus requires to take into account the mutual interactions of  IGW and dynamo-generated magnetic field. Such effects depend on stellar conditions and can evolve along with their structural evolution.

\tck{We model the dynamics induced by IGW and dynamo in rotating radiative stellar layers using a simple description applicable to various evolutionary stages. As dynamo action and the propagation of IGW are 3D processes that have characteristic timescales very short compared to typical periods associated with structural evolution of stars, we \tck{propose} a mean-field 1D model by taking advantage of the dynamo coefficients computed from 3D spherical simulations. In this model, the necessary mean shear flow to trigger the dynamo results from the dissipation of monochromatic IGW generated in existing adjacent convective layers, which are \tck{expected} to drive the formation of an oscillating rotational shear layer, the so-called Shear Layer Oscillation (SLO). In turn, magnetic effects can act on the mean flow through the Lorentz force.}

\tck{We show that the inclusion of magnetic fields adds up to the already very complex nonlinear problem and} gives rise to the emergence of new dynamical regimes. \tck{Particularly, the fast SLO generated very close to the place where IGW are generated is perturbed by magnetic fields. This dynamical change} can filter the wave energy spectrum transmitted towards \tck{further layers, with potential influence on the long-term evolution of the inner rotation.} Our work constitutes a first attempt to account for IGW and magnetic fields interacting dynamically through Lorentz forces and a mean flow in a 1D \tck{dynamo} model of interest for studying long-time evolution of stars.
  

\end{abstract}

\section{Introduction}

Studying the impact of turbulence and global rotation in stably stratified layers, \tck{as well as} their interplay with magnetic fields, is ubiquitous in many research areas addressing fundamental questions related to the Earth's biosphere, atmospheric and oceanic flows as well as stellar interiors. 
In the last twenty years, space-borne helio- and asteroseismic missions such as SoHO \citep{Domingo1995}, CoRoT \citep{Baglin2006} and {\it Kepler} \citep{Borucki2010} enabled us to unveil some key physical properties inside stars but also highlighted our inability to properly describe the complex dynamics in their \tck{radiative stably stratified} interiors. A striking example is our incapacity to model the angular momentum (AM) redistribution in the Sun and low-mass stars \citep[e.g.][]{AertsMR2019}. This is currently a stumbling block as it impacts the determination of stellar fundamental parameters such as the age and subsequently our knowledge of the structure and evolution of planetary systems as well as the chemodynamical history of the Milky Way \citep[e.g.][]{Miglio2017}. The redistribution of AM in stellar stably stratified regions is expected to be driven by diverse processes, such as stellar winds, structural modifications (contraction and dilatation), meridional circulation, waves (generated at the interfaces between convective and radiative regions), turbulence, and magnetic field, but no clear picture yet emerges on their respective role and interplay \citep[see for instance][]{Maeder2009}. 
\tck{In this context, we focus on AM transport due to internal gravity waves (IGW, hereafter) and magnetic fields, as these two processes appear as promising candidates to shape rotation profiles in stars.}

On the one hand, the potential of IGW to transport angular momentum has been studied for decades in stellar physics \citep[e.g.][]{Press81,Schatzman1996}. These waves, for which buoyancy serves as the restoring force, are generated by turbulent motions at the edges of convective regions of stars and propagate into the adjacent radiative layers. They can notably be damped by radiative diffusion and deposit as well as extract angular momentum into the medium, which modifies locally the rotation profile \citep[e.g.][]{zahn97,Kumar99}. 
While rapidly recognised to be able to explain the quasi-solid rotation in solar radiative zone \citep{TalonC05}, recent studies also indicate that IGW could potentially slow down the core rotation of more evolved stars on the subgiant and red giant branches \citep{Fuller2014,BelkacemMGOMM15,PinconBGM17}, \tck{as well as explain the differential rotation of massive main-sequence stars \citep{Rogers2015}}. However, these preliminary estimates call for more sophisticated investigations as several important simplifying assumptions, related to the complex interplay between the waves and the rotation, have been made. 
In particular, \tck{they did not properly take into account} the wave-driven formation of an oscillating rotational shear layer just below the base of the convective zone, the so-called Shear Layer Oscillation (SLO). This is equivalent to the Quasi-Biennal Oscillation (QBO) \tck{which generates equatorial periodic winds} in the Earth's atmosphere, \tck{evolving on a period of about 28 months} \citep{baldwin01}, or the Quasi-Quadrennial Oscillation for Jupiter \citep{Leovy1991}, for which the period is approximately 54 months. The SLO can filter the wave energy spectrum transmitted towards deeper layers and can therefore influence the long-term evolution of the inner rotation \citep{TalonKZ02}. The nonlinear dynamics of such a SLO (or QBO) has been studied in several works, either based on the 1D model by \cite{Plumb1978} initially developed for the QBO \citep[e.g.,][]{KimM01,Renaud2019,Renaud2020,Leard2020b}, laboratory experiments \citep[e.g.,][]{semin16,semin2018,Leard2021} and \tck{Cartesian }direct numerical simulations (DNS) \citep[e.g.,][]{Couston2018}. \tck{However the possibility of a SLO in a stellar context is still debated, \tck{notably due to the many uncertainties concerning for instance the period of the oscillation, estimated to vary according to the solar cycle \citep{Kumar1999} or on a few hundreds years \citep{TalonKZ02} for the Sun, or on few years for stars slightly heavier than the Sun \citep{Talon2003}.} Previous numerical simulations of full stars in 2D \citep[e.g.,][]{Rogers2006} or 3D \citep[e.g.,][]{Alvan2014} in spherical geometry have not reproduced such an oscillation, \tck{possibly because they evolve in regimes with unrealistically high-diffusivity coefficients. In a forthcoming paper \citep{Daniel2025}, we nevertheless demonstrate the possibility of such a phenomenon in 2D polar geometry when thermal diffusion dominates over kinematic viscosity. Such simulations are however very demanding in terms of computational power}. In this regard, due to the intrinsic limitation of global DNS, reduced models offer a first interesting perspective to study the effects of IGW on angular momentum transport \citep{Plumb1978, RenaudNadeau2019}. Although very simplified compared to the entire dynamics of a star, they enable a first approach to understand how the mean flow and the waves can interact to influence AM transport in a stellar environment.}

On the other hand, magnetic fields are also thought to play a major role in the redistribution of AM in stars \citep[e.g.,][]{EggenbergerDB19}. Strong magnetic fields can significantly affect the dynamics of a plasma flow, inducing magnetic braking and potentially triggering magnetohydrodynamic (MHD) turbulence---thus enhancing both effective viscosity and angular momentum transport.
The recent rise of ground-based spectropolarimeters (e.g., NARVAL, ESPaDOnS, SPIRou) has already provided surface estimates of stellar magnetic fields magnitudes and topologies \citep[e.g.,][]{donati2023.525.2015D} and the space-borne mission {\it Kepler} has permitted for the first time to directly infer internal magnetic fields in radiative interiors \citep{LiDBL2022,Li2023,Deheuvels2023}. 
However, despite the growing amount of available constraints, understanding the origin of magnetic fields, their diversity, and their feedback on stellar plasmas remains a considerable theoretical challenge. 
In this context, \tck{and related to the SLO, the effect of adding a magnetic field on the dynamics of IGW has been investigated \citep{KimM03}, showing a strong modification of the oscillating layer and thus of the AM transport properties when a toroidal field is applied \citep{Rogers2010,Macgregor2011} or when were both simulated a radiative and a convective zone \citep{Rogers2011}.} Magnetic instabilities have \tck{also} been intensively investigated, in particular the magneto-rotational instabilities \citep[e.g.,][]{petitdemange13,Rudiger2015,Meduri2024} and the Tayler-Spruit dynamo \citep[e.g.][]{Fuller19,Eggenberger2022}. The latter was initially proposed by \citet{Spruit02} and is based on a dynamo mechanism occurring in radiative stably stratified regions. In this scenario, the magnetic field generation relies on the destabilisation of a strong, toroidal and axisymmetric magnetic field by Tayler instability \citep{tayler73}, which creates a poloidal field and thus (in theory) closes the dynamo loop initiated by differential rotation. \tck{As for the SLO}, the existence of this Tayler-Spruit dynamo has been debated for a long time \citep{Braithwaite2006,ZahnBM07} as well as its ability to explain observations \citep[e.g.][]{Cantiello2014,Fuller19,Eggenberger2022}.
\tck{If recent simulations by \citet{PetitdemangeMG23,DanielPG23} and \citet{Petitdemange24} have proven the possibility of a highly-efficient dynamo to transport angular momentum in radiative zones, the highly-idealised setup they considered where a large scale shear is imposed (see section \ref{alphaDNS}) can raise some reasonable concern as to the relevance for astrophysical applications. In addition, while their results showed some similarities with Spruit's formalism, other aspects of the theory were not verified, particularly about the location of the saturated field, situated close to the equator, while one would rather expect the field to be near the poles \citep{Spruit99}. Nevertheless, the fact that a dynamo is obtained in a radiative zone, and that it is highly subcritical, i.e. that it can be maintained at a very low level of shear \citep{DanielPG23} makes it an interesting possibility for stellar interiors where waves can provide the necessary shear, as small as it may be compared to the rotation of the star.}

The scope of this work is thus to build on the recent 3D DNS results of \citet{PetitdemangeMG23,DanielPG23} and \citet{Petitdemange24} to investigate the mutual interaction between the obtained dynamo and IGW. Indeed, these results brought some numerical evidence that contrary to past theoretical studies \citep{Goossens1981, Spruit99}, a dynamo could occur in regions of mid latitudes or centred on the equator, quite similar to the Tayler-Spruit’s (TS) one. In these regions the wave-induced mean flow is expected to develop. Our approach consists in studying the overlap of the two phenomena through a reduced model in which the differential rotation needed to trigger Tayler instability originates from the wave-induced oscillating mean flow. As a first step in the investigation, we propose a simple new 1D description. Such 1D models can serve as a guideline for future numerical investigations of the problem through global simulations. Moreover, they can also be of interest to go beyond the usual picture regarding the effect of the magnetic fields and waves on the angular momentum redistribution in 1D stellar evolution codes, which mostly rely on simplistic theoretical arguments (e.g., diffusive modelling) and neglect interaction between both \citep[e.g.,][]{Spruit02,Maeder2003,Cantiello2014,Fuller19}. \\

Adopting a separation of scales between small fluctuations and average background quantities, we first use the recent set of simulations by \citet{PetitdemangeMG23,DanielPG23} and \citet{Petitdemange24} to measure the mean-field coefficients of the dynamo. We also recall more in details the similarities and discrepancies of this previous work with TS dynamo, highlighting the idealisation of our simple model (see section \ref{alphaDNS}). Second, considering a thin local box in the radiative zone, we add the effect of monochromatic IGW generated at a convective boundary in the mean field equations (section \ref{section_modele_1d}). The wave-induced oscillation therefore represents the natural source of differential rotation required to trigger the dynamo, in contrast with previous numerical simulations where the shear results from an imposed Couette flow. In this spirit, our new local 1D model can be seen as a MHD version of the hydrodynamical model of \citet{Plumb1978}, including the induction equation with a parameterisation of the electromotive force resulting from the Tayler-Spruit dynamo. This toy model also proves to be a new setup for the study of new dynamical regimes as a modification of the already very rich nonlinear model of \citet{Plumb1978}. Many studies have focused on purely hydrodynamical models in the past on a nonlinear perspective and we give here some non exhaustive examples of how these results can be modified when the flow is coupled to the magnetic field (section \ref{section_results}). Finally, we discuss astrophysical applications of our approach and the prospects for extrapolating to realistic stellar regimes (section \ref{section_discussion}).

\section{From 3D spherical simulations towards a simple 1D modelling}
\label{method}

\tck{In this section, we expose the strategy we adopt to build our 1D toy model and the underlying assumptions. We first remind the results obtained from the recent 3D simulations and how they are used to characterise the \tck{obtained} dynamo in a simple way through the measure of the mean-field $\ds \alpha$-effect (subsection \ref{alphaDNS}). We then integrate this parameterisation of the dynamo into the mean flow equations considering a thin local box and monochromatic IGW emitted at one of its boundary (subsection \ref{section_modele_1d}).}

\subsection{3D DNS approach}
\label{alphaDNS}

\subsubsection{Numerical setup}
As previously described in \citet{PetitdemangeMG23}, a radiative envelope is modelled as a spherical shell of inner and outer radii $\ds r_i$ and $\ds r_o$, the spheres rotating respectively at $\ds \Omega_o + \Delta \Omega$ and $\ds \Omega_o$, with a fixed aspect ratio $\ds \chi=r_i/r_o=0.35$. The flow is considered incompressible and we use Boussinesq approximation, i.e. the density variations are only taken into account in the buoyancy term. In this approximation, the stably stratified nature of the radiative zone is achieved by applying a temperature gradient between the two spheres $\ds \Delta T= T_o-T_i>0$ \citep[e.g.][]{Gage1968}. In all the 3D simulations analysed in this paper, we used the pseudo-spectral code PaRoDy code \citep{dormy98,Aubert2008} coupled to the ShtNs library \citep{Schaeffer2013} to integrate the magnetohydrodynamics equations for the flow velocity $\ds \bf v$, the magnetic field $\ds \bf B$, the pressure $\ds P$ and the temperature perturbation $\ds \Theta$ taken to be the deviation from the stationary solution \citep{DanielPG23}. The integrated equations are the following:
\begin{align}
\nonumber   &\frac{\partial \mathbf{v}}{\partial t} + (\mathbf{v}\cdot\bm{\nabla})\mathbf{v}  = -\bm{\nabla}P  - 2 \frac{ 1}{EkR_e} \mathbf{e_z} \times \mathbf{v} + \frac{1}{R_e}\bm{\Delta} \mathbf{v} \\ 
    &+ \left(\frac{1}{EkR_e} + \frac{1}{\chi}\right)(\bm{\nabla} \times \mathbf{B})\times \mathbf{B} 
    +\frac{Ra}{PrRe^2} \Theta (r/r_o) \mathbf{e_r},  \label{3:eq:EQ_TS_ADIM_1} \\
    &\nabla\cdot \mathbf{v} =0, \label{3:eq:EQ_TS_ADIM_2}\\
    &\frac{\partial \Theta}{\partial t} + (\mathbf{v}\cdot\bm{\nabla})\Theta = \frac{1}{PrRe} \Delta \Theta - \frac{\chi}{1-\chi}\frac{v_r}{r^2},\label{3:eq:EQ_TS_ADIM_3}\\
    &\frac{\partial \mathbf{B}}{\partial t} = \bm{\nabla} \times \left(\mathbf{v} \times \mathbf{B} \right) + \frac{1}{PmRe} \bm{\Delta} \mathbf{B}, \label{3:eq:EQ_TS_ADIM_4}\\
    &\nabla\cdot \mathbf{B} =0, \label{3:eq:EQ_TS_ADIM_5}
\end{align}
\label{3:eq:EQ_TS_ADIM}
\noindent\ignorespacesafterend where the dimensonless parameters $Ek=\nu / (r_o^2\Omega_o)$, $Re=r_o r_i\Delta \Omega / \nu$, $ Ra=\alpha_T \Delta Tg_o r_o^3/(\nu\kappa)$, $Pr=\nu / \kappa$ and $ Pm=\nu / \eta$ are respectively the Ekman, Reynolds, Rayleigh, Prandtl and magnetic Prandtl numbers. In their definitions, $\ds \nu,\kappa$ and $\ds \eta$ are respectively the flow diffusion coefficient, and the thermal and magnetic diffusivities, which are taken to be constant throughout each simulation. $\ds \alpha_T$ is the thermal expansion coefficient, and $\ds g_o$ the value of the gravity acceleration at $\ds r=r_o$, whose radial profile is considered linear, $\ds g = g_o(r/r_o)$. The amplitude of the stable stratification can be described by a global Brunt-Väisälä frequency $\ds N = \sqrt{\alpha_T g_o \Delta T/(r_o-r_i)}$. All the dimensionless parameters were varied as reported in the simulations described in \citet{PetitdemangeMG23} and \citet{DanielPG23}, that is, $\ds Ek \in [10^{-7},10^{-4}]$, $\ds Re \in [3.10^2,10^5]$, $\ds Ra \in [10^8,10^{10}]$, $\ds Pm \in [0.5,25]$. $\ds Pr$ was fixed to $\ds 0.1$. 

\tck{Anticipating on the following section where the 1D model will be presented, we briefly recall how the dynamo was obtained in the 3D simulations, \tck{and the similarities it shares with Spruit's picture.}}

\subsubsection{Past results \tck{and dynamo regime}}
\label{section_discussion_TS}

From a nonlinear physics point of view, the model described here corresponds to a spherical Couette problem, host to a full range of rich and complex behaviours depending on the values of the parameters considered and which has been studied in detail in the past in some limits: see for instance \citet{Hollerbach06} or \citet{Wicht2014} for a purely unstratified hydrodynamical study, and \citet{guervillyC10} or \citet{Gissinger2011} to see the effect of the magnetic field.
\tck{Adding stratification complicates the physics at hand, and several different dynamo scenarios have been envisaged inside such environments. \tck{Besides} \citet{Spruit99} \tck{predicting} that the Tayler instability would be the more likely to develop, the possibility of the azimuthal magneto-rotational instability has also been investigated \citep[e.g.,][]{Meduri2024}. In this context, recent results by \citet{PetitdemangeMG23, Petitdemange24} have brought some attention as they showed many similarities with the formalism of Spruit, but failed to verify other important steps predicted by the theory \citep[e.g.,][]{Ma2019}. As the matter is at the core of the toy-model developed in section \ref{section_modele_1d}, and because the question has received recent theoretical development \citep{Skoutnev2024}, we remind here the key aspects of the theory used to build our model, whose use is motivated by 3D DNS results.}

\tck{Spruit makes use of the so-called shellular assumption stating that the angular velocity $\ds \Omega$ is of spherical symmetry due to the important stratification \citep{zahn92}. The form of an initial toroidal field is assumed to be $\ds B_T\propto s^p \cos \theta$, with $\ds s$
the distance from the axis of rotation and $\theta$ the polar angle. As recalled by \citet{Skoutnev2024}, $p=1$ is the most general case if there is any current density on the axis, but $p$ is kept as a free parameter characterising the toroidal field for the following local approach, as many forms can be envisaged in a fully turbulent stellar environment. Near the rotation pole of the star where $\ds \cos \theta \sim 1$, the previous form of the field simplifies. This \textit{a priori} necessary condition imposes to be close to the pole but it can be by-passed if the preliminary/initial field already exhibits locally such a radial dependency; it is the case in DNS of a weak-field dynamo branch (see \citet{PetitdemangeMG23} and after).} Considering the general case of a rotating star whose diffusion cannot be neglected, \citet{tayler73} and \citet{Spruit02} have derived the following necessary local conditions\tck{---meaning that they must be verified somewhere in the flow---}for the \tck{Tayler} instability:
\begin{align}
    & p>\frac{m^2}{2} -1 \quad (m\neq0), \quad p>1 \quad (m=0), \label{eq:criteres A}\\
    & \frac{\omega_A}{\Omega} > \left(\frac{N}{\Omega}\right)^{1/2}\left(\frac{\eta}{\kappa}\right)^{1/4}\left(\frac{\eta}{r^2 \Omega}\right)^{1/4}.
    \label{eq:criteres TI}
\end{align}
where $\ds m$ is the azimuthal order of the considered mode, $\ds \omega_A= B_\phi/\sqrt{\rho \mu_0 r^2}$ and $\ds N$ are the Alfvén and  Brunt-Väisälä frequencies, respectively.

\tck{As presented in \cite{PetitdemangeMG23}, the resulting field in the DNS is a dominant large-scale azimuthal one, close to the equator, for which \tck{Eqs.~(\ref{eq:criteres A}-\ref{eq:criteres TI})} are well verified. Indeed, criterion (\ref{eq:criteres A}) can be translated into $\ds Rb>0$, where $\ds Rb:=(p-1)/2$ is a parameter introduced by \citet{Kirillov2014}, who showed that the Tayler instability would dominate in this case over the magneto-rotational one. This holds in our simulations in the region where the magnetic field is the strongest, i.e. at the equator \citep{Petitdemange24}. The second criterion (\ref{eq:criteres TI}) can also be checked against in the simulations. \citet{PetitdemangeMG23} pointed out that when the fiducial simulation they reported saw its magnetic energy cross a threshold in term of $\ds \Lambda = (\omega_A/\Omega)^2(Pm/Ek)>(N/\Omega)(Pr/Ek)^{1/2}$, the field would become unstable. As explained in their supplementary material, this condition is a \textit{local} one. Interestingly, the thus obtained azimuthal field is much stronger that the radial one. The latter is associated to a length scale, namely the Tayler scale $\ds l_c$ defined as: }

\begin{equation}
    l_c = r\left(\frac{\Omega_o}{N}\right)^{1/2}\left(\frac{\kappa}{r_o^2\Omega_o}\right)^{1/4},
    \label{lc_spruit}
\end{equation}

\noindent \tck{to which can be related the ratio $\ds B_r/B_\phi \sim l_c/r$. This is somehow true in our case, at least in orders of magnitude \citep{Petitdemange24}. To this general agreement between \citet{Spruit02} and our simulations\tck{---radial profile and amplitude of $B_\phi$, as well as scale separation with $B_r$---}must be added the behaviour of the Maxwell stress \citep{PetitdemangeMG23, Petitdemange24} and the minimum level of differential rotation needed to maintain the dynamo \citep{DanielPG23}, both remarkably close to theoretical estimates.} 

\tck{It is nevertheless as important to mention the points that fail to agree with the theory. The most crucial one seems to be the location of the dynamo, situated at the equator and not at the poles. The latitudinal dependency of the Tayler instability has been notably investigated by \citet{Goossens1981} who showed that the instability was more likely to occur close to the rotation axis in an idealised adiabatic system, which was confirmed more recently by \citet{Ma2019}}. 
\tck{As recalled by \citet{Skoutnev2024}, previous studies have considered idealised and simplified angular dependency for the stability of the toroidal field, that is $\ds B_T \propto \sin \theta$ or $\ds B_T \propto \sin\theta \cos \theta$, corresponding to the marginal case for stability of the toroidal field ($\ds p=1$). It may indeed be possible that a different latitudinal geometry for the field may affect this traditional Spruit's picture occuring close to the axis, as stronger values of $\ds p$ could affect its stability.}
\tck{
The question of the precise localisation of the magnetic instability therefore appears as a discrepancy between DNS and theory. It suggests that the generation of the field we observe \tck{in the DNS} might only be achieved through some meridional transport to comply with the \textit{classical} picture of the TS dynamo. The previous discussion, besides being a reminder of past studies, emphasises that direct comparisons between simulations and theory are sometimes made difficult due to a different set of assumptions in the two approaches; in our DNS, the flow is not shellular and the dominant toroidal field cannot be described by a simple $\ds \sin \theta$ latitudinal dependency (see figure \ref{fig_alpha_measure}($\ds a$)). Nevertheless in the following, the precise localisation of the instability does not matter as, inspired from DNS results, we adopt a local approximation to constrain nonlinear effects.}

\tck{Beyond the obvious differences between local analysis and global simulations, one should keep in mind the key question at hand here, which is the transport of angular momentum in a radiative stellar context. The results described in \cite{PetitdemangeMG23, Petitdemange24} exhibit a mechanism which can extract angular momentum, through the nonlinear coupling of Maxwell and Reynolds stresses, and that can be maintained at arbitrary low differential rotation as long as the induction present in the plasma is high enough \citep{DanielPG23}. The controversy on whether this mechanism should be depicted as Tayler or not is actually secondary on the question of AM transport. This is why in what follows we use the many similarities with the TS dynamo to build our model, but the reader should keep in mind the above discussion: the TS formalism is a pretext to show how the properties of the wave-induced mean flow are modified with a simple parameterisation of dynamo action.}

\tck{In this regard, we particularly make use of the strong scale separation between the radial and azimuthal components of the magnetic field}, which naturally suggests the introduction of a mean field formalism \tck{as proposed in the following.} \tck{In particular, based on the simulations in \citet{DanielPG23}, we know that in the mean-field paradigm, the dynamo can be characterised through the so-called $\ds \alpha$-effect.}

\subsubsection{A measure of the $\alpha$-effect in Direct Numerical Simulations}

\tck{We decompose any physical quantity $\ds X$ as $\ds X = \overline{X} + X^*$ with $\ds \overline{X^*} =0$, where we have defined the azimuthal mean:
\begin{equation}
\ds \overline{X} = \ds \frac{1}{2\pi}\int_{0}^{2\pi} X d\phi \;.
\end{equation}
To go further, one can average Eq.~(\ref{3:eq:EQ_TS_ADIM_4}) to obtain \citep{moffatt}:}
\begin{equation}
    \frac{\partial{\bf \overline{B}}}{\partial t} = \frac{1}{RePm}\Delta{\bf \overline{B}} + {\bf \nabla}\times({\bf \overline{v}}\times{\bf \overline{B}}) + {\bf \nabla}\times {\boldsymbol{\mathcal{E}}},
    \label{Indmean}
\end{equation}

\noindent where $\ds {\boldsymbol{\mathcal{E}}}=\overline{{\bf v}^*\times {\bf b^*}}$ is the electromotive force (EMF), with $\ds {\bf v^*},{\bf b^*}$ the perturbations from the averages of $\ds \bf v$ and $\ds {\bf B}$, respectively. Making use of the linearity of Eq.~(\ref{Indmean}) in $\ds \overline{\bf B}$ one can get the following expansion: 
\begin{equation}
\mathcal{E}_i = a_{ij} B_{0j} + ...,
\label{mean field}
\end{equation}
where Einstein notations have been used with summation over repeated indices ($\ds i=r$, $\ds \theta$ or $\ds \phi$) and with $\ds a_{ij}$ the coefficients of  $\ds {\bf a}$ an order 2 tensor describing the correlations between the EMF and the mean field $\ds {\bf B_0} := \overline{{\bf B}}$. Note that for simplification we limit ourselves to a first order expansion but higher-order terms in spatial derivatives could be considered. Splitting the \textcolor{black}{mean} fields into toroidal \textcolor{black}{(${\bf B_T}, {\bf v_T}$)} and poloidal \textcolor{black}{(${\bf B_P}, {\bf v_P}$)} parts, Eq.~(\ref{Indmean}) can be projected as \citep{raedler80}:
\begin{align}
\frac{{\rm D} {\bf B_T}}{{\rm D} t} &=   \frac{1}{RePm}\Delta {\bf B_T} + {\bf B_P}\cdot {\bf \nabla v_T} + {\bf \nabla} \times (\bf a B_P), \label{toro}\\
\frac{{\rm D} {\bf B_P}}{{\rm D} t} &=   \frac{1}{RePm}\Delta {\bf B_P} + {\bf \nabla} \times ( \bf a B_T),\label{polo}
\end{align}

\noindent where $\ds {\rm D}/\rm Dt  = \partial /\partial t + {\bf v_P}\cdot{\bf \nabla}$ stands for the material derivative and $\ds \bf a B_T$ the product of $\ds \bf a$ by $\ds \bf B_T$.

Several regimes can be distinguished then, depending on which source term dominates in Eq.~(\ref{toro}). Indeed, if $\ds \bf B_P$ can only be amplified by $\ds \alpha-$effect---$\ds {\bf \nabla} \times (\bf a B_T)$---$\ds \bf B_T$ can either be amplified by $\ds \Omega-$effect, corresponding to $\ds {\bf B_P}\cdot {\bf \nabla v_T}$, or an $\ds \alpha$ term $\ds {\bf \nabla} \times ( \bf a B_P)$, describing classical $\ds \alpha-\Omega$ or $\ds \alpha^2$ dynamo solutions, or even $\ds \alpha^2-\Omega$ if the last two terms in Eq.~(\ref{toro}) are \textcolor{black}{of comparable magnitudes}. Following \citet{DanielPG23}, in which Tayler-Spruit dynamos were described as compatible with $\ds \alpha-\Omega$ solutions, the second source term in Eq.~(\ref{toro}) will be dropped in what follows.

Introducing $\ds \boldsymbol{\alpha}$ and $\ds \boldsymbol{\gamma}$ the symmetric and antisymmetric parts of $\bf a$, one can rewrite Eq.~(\ref{mean field}) as:
\begin{equation}
    \mathcal{E}_i = \alpha_{ij}B_{0j} + \left(\boldsymbol{\gamma}\times \bf B_0\right)_i.
\end{equation}
The tensor $\ds \boldsymbol{\alpha}$ is usually seen as a generalisation of $\ds \bf a$, which is going to play the role of a source term in Eq.~(\ref{toro}) and (\ref{polo}), while $\ds \boldsymbol \gamma$ plays the role of an effective velocity advecting the mean magnetic field \citep{schrinner07,schrinner12}. Their components are related to $\ds \bf a$'s as:
\begin{equation}
    \alpha_{ij} = \frac{1}{2}\left(a_{ij} + a_{ji} \right), \quad \gamma_k = \frac{1}{2}\epsilon_{kij}a_{ij}.
\end{equation}

\tck{In order to be able to implement a simple $\ds \alpha-\Omega$ 1D model with little computational cost, we measured the coefficients of the tensor $\ds \bf a$ in our data set of simulations. Several methods exist to measure such mean-field components as for instance the test-field method \citep{schrinner07}. As \tck{the test-field method} is rather heavy to implement in 3D simulations, we chose to work with the Singular Value Decomposition (SVD) approach \citep{Racine2011,Simard2016,Dhang2020}.  }
The idea of this approach is to bypass the \textit{a priori} lack of equations in Eq.~(\ref{mean field})---3 equations for 9 $\ds a_{ij}$ unknown---through a minimisation process. Details are given in Appendix~\ref{SVD}, in which we compare the SVD approach to the test-field method for a run reported in \citet{schrinner12} \tck{and show the results are comparable }.
\\

\begin{figure}[t]
\begin{center}
\begin{minipage}{150mm} 
\subfigure[]{
\resizebox*{7.5cm}{!}{\includegraphics{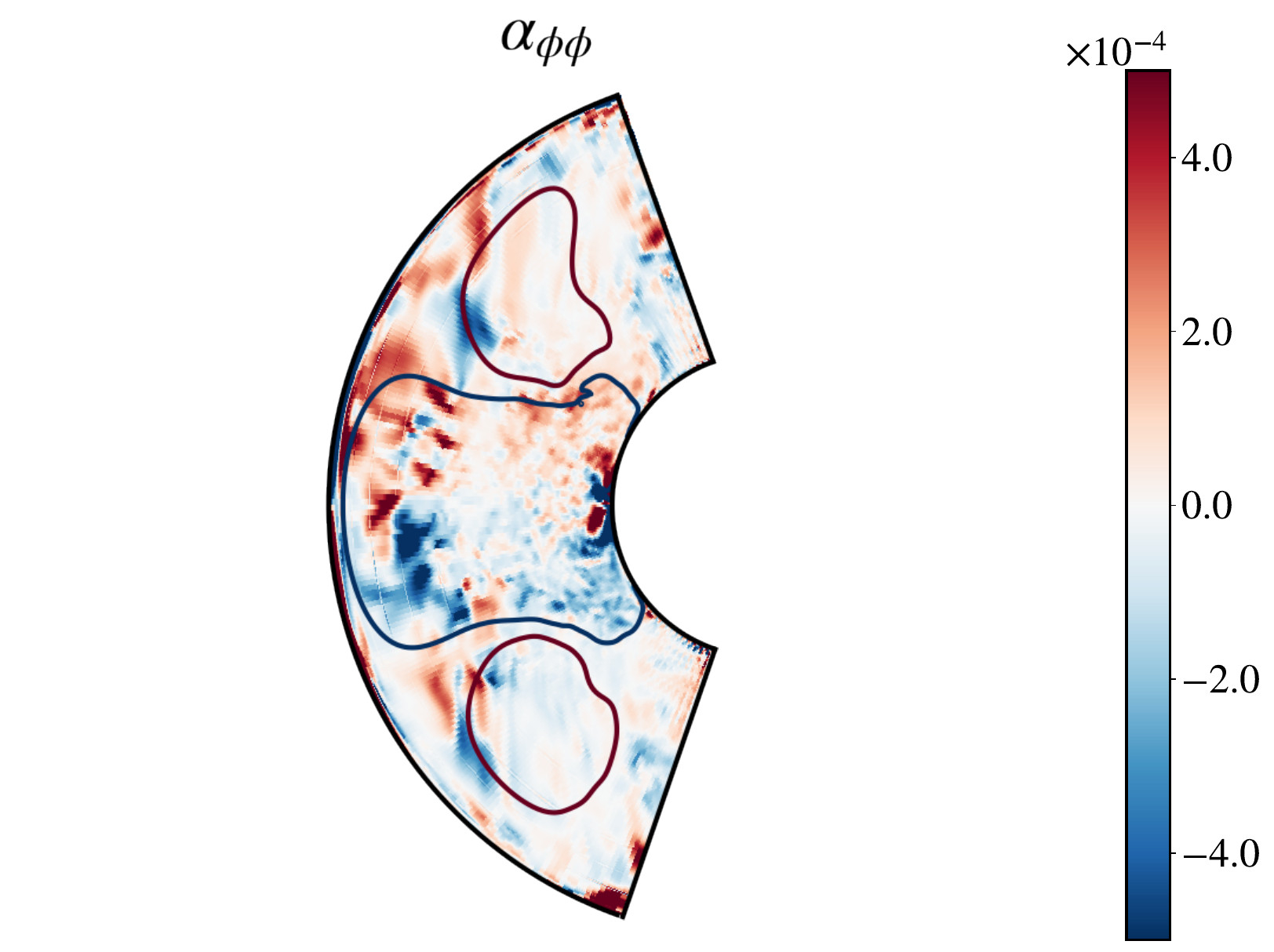}}}%
\subfigure[]{
\resizebox*{7.5cm}{!}
{\includegraphics{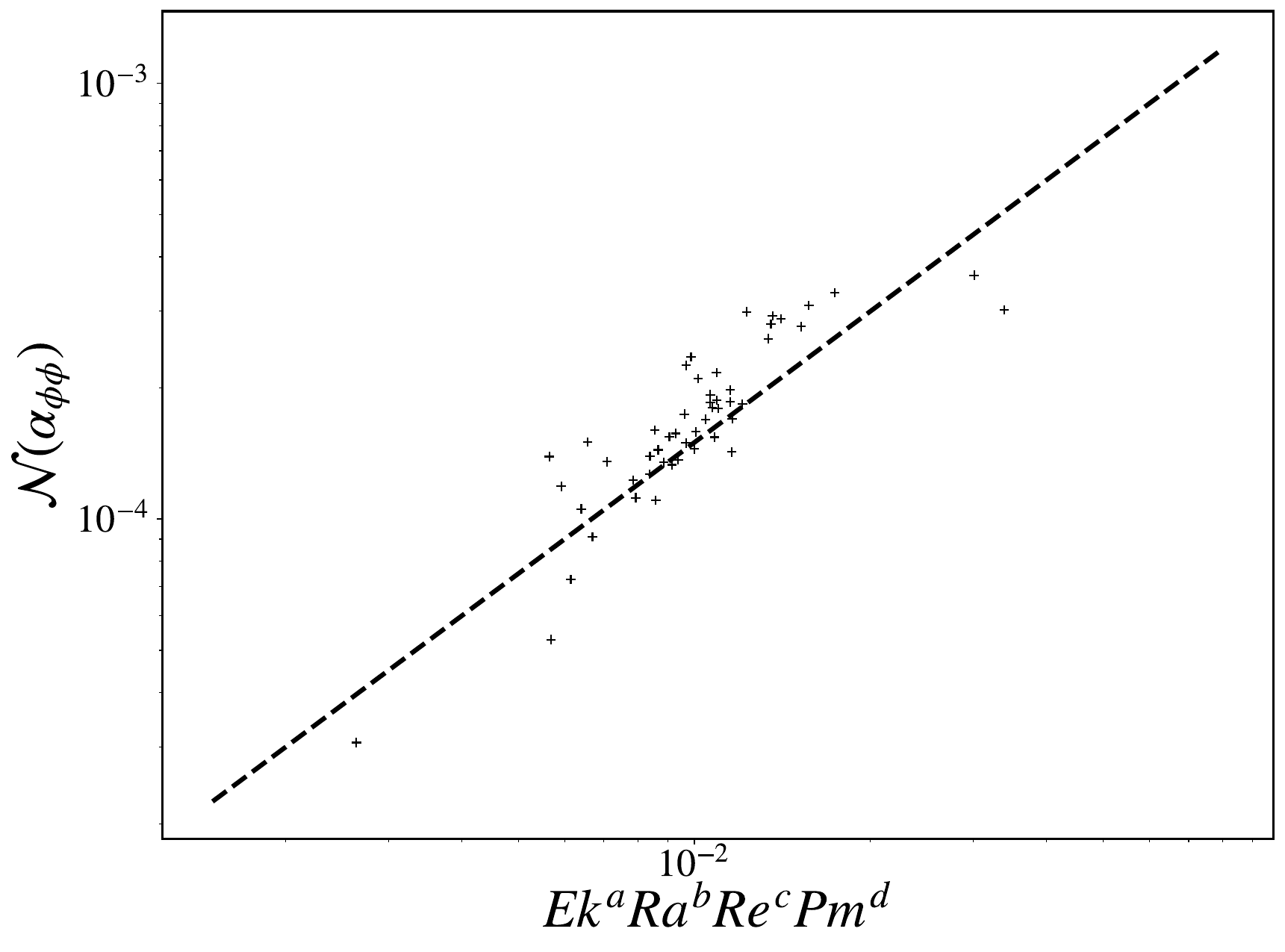}}}
\caption{A measure of the $ \alpha$-effect. \textit{(a)} Meridional section of the component $\ds \alpha_{\phi\phi}$ measured with the SVD technique applied to the simulation $\ds Ek=10^{-5}$, $\ds Ra=10^9$, $\ds Re=27500$, $\ds Pr=0.1$ and $\ds Pm=1$ \citep{PetitdemangeMG23}. Isovalues of the time-averaged azimuthal magnetic field $B_\phi$ are visible. \textit{(b)} Linear regression of $\ds \mathcal{N}(\alpha_{\phi\phi})$ for our data set.}%
\label{fig_alpha_measure}
\end{minipage}
\end{center}
\end{figure}

Making use of this technique for our data set, one can get a map \tck{of each coefficient} as the one displayed in figure~\ref{fig_alpha_measure}($\ds a$) for $\ds \alpha_{\phi\phi}$. \tck{$\ds \alpha_{\phi\phi}$} is going to be the most significant one as the dominant part of the magnetic field is its azimuthal component, which is centred around the equator (see some isovalues on figure \ref{fig_alpha_measure}($\ds a$)). \tck{Note that this is characteristic of our 3D results. Indeed, a different setup studied by \citet{Barrere2023} also exhibits many similarities with \citet{Spruit02}'s formalism but leads to a dynamo located close to the poles.} 

In our case we see on figure \ref{fig_alpha_measure}($\ds a$) that the $\ds \alpha_{\phi\phi}$ obtained is indeed very important at the equator, as it is a measure of the correlation of the EMF and the magnetic field. It remains however poorly captured close to the poles. Thus, we will focus on the strong field region for extracting a relevant value  used for parameterising the $\alpha$-effect. \tck{To do so, we need a way of quantifying $\ds \alpha_{\phi\phi}$ for each parameter $\ds Ek, Ra..$, going from a 2D $\ds (r,\theta)$ map to a scalar. We therefore require a norm which we choose to be:}
\begin{equation}
	\mathcal{N}(\alpha_{ij}) = \sqrt{\max_{r\in [0.4,0.7]} \int_{\pi/4}^{3\pi/4} \alpha_{ij}^2(r,\theta) \sin\theta d\theta},
	\label{norme}
\end{equation}
and study how it varies with the control parameters. The norm (\ref{norme}) is \tck{a choice}. The motivation is to focus on the zone where the dynamo effect considered is dominant, i.e. centred around the equator, so as to filter out the non-relevant values close to the poles. Other filters could have been considered \citep[see for instance][]{Livermore2010}.

We report in figure \ref{fig_alpha_measure}($\ds b$) the result of a linear regression in which the dependency of $\ds \mathcal{N}(\alpha_{\phi\phi})$ (in code units, \tck{that is in units of $r_i\Delta\Omega$}) is searched as proportional to $\ds Ek^a Ra^b Re^c Pm^d$, which leads to $\ds a=-0.07$, $\ds b=-0.42$, $\ds c=0.35$ and $\ds d=0.17$. Caution need be taken when considering this fit. Indeed, as it appears on figure \ref{fig_alpha_measure}($\ds b$), the measure presented here can be very dispersed. However, it enables us to get some estimate of the variation of the coefficient as a function of the parameters \tck{so as to be able to implement this estimate in a 1D code}. Such a dependency could be refined in a future study. \tck{We nevertheless get a behaviour of $\ds \alpha_{\phi\phi}$ with the parameters which makes sense regarding the respective signs of the previous exponents; an increased rotation leads to stabilising effects and a decrease of $\ds \alpha_{\phi\phi}$, and the same can be said about stratification and the Rayleigh number. On the contrary, increasing large-scale shear through the Reynolds number leads to a larger value of $\ds \alpha_{\phi\phi}$.}

\subsection{1D model of the magnetised wave-induced oscillation}
\label{section_modele_1d}

\tck{As recalled in the introduction, IGWs excited in convective regions can propagate in the adjacent radiative layers and \tck{possibly} generate a mean shear oscillation induced by the Reynolds stresses. Several works modelled this phenomenon in an hydrodynamical way \citep{Plumb1978, Renaud2019}, without accounting for the possibility of a Tayler-Spruit dynamo to set in due to the strong shear gradient. In this section, we aim to build a simple 1D model including the generation of magnetic fields by TS dynamo and its feedback on the mean flow by Maxwell stresses and relying on the parameterisation based on previous DNS. \tck{Note that contrary to some past studies who investigated the role of an imposed magnetic field on the dynamics of IGW and mean flow oscillation \citep[e.g.][]{Rogers2010, Macgregor2011}, our setup allows for the field to grow and maintain itself in time through dynamo action because of the parameterisation of the $\ds \alpha-$effect we use. We therefore expect significant differences with these past studies who notably observed a time-independent shear layer or wave reflection due to the strong magnetic field applied.}}

\begin{figure}[t]
\begin{center}
\begin{minipage}{120mm}
{\resizebox*{12cm}{!}{\includegraphics{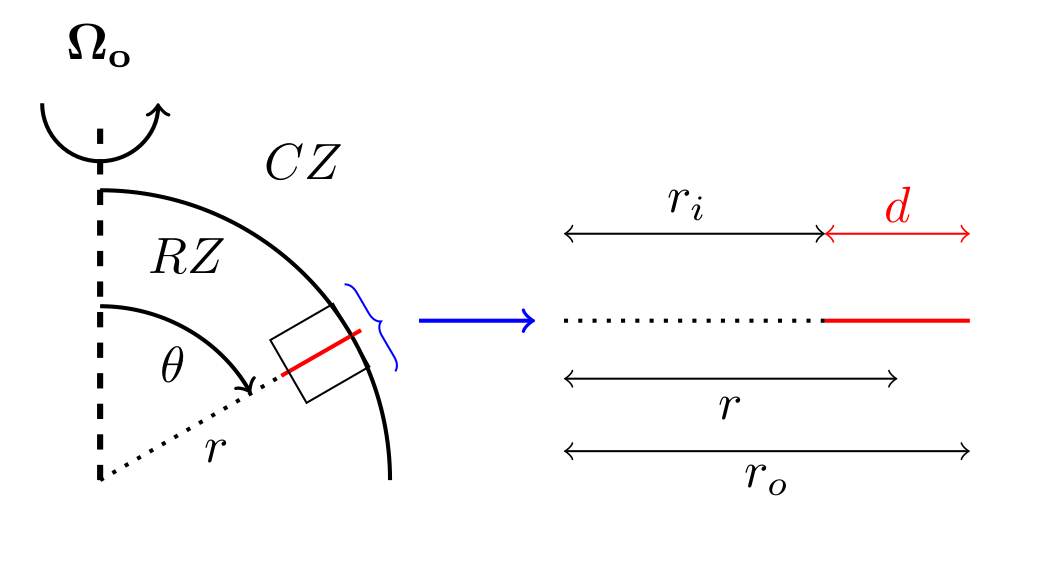}}}%
\caption{Geometry of the domain (in red) (not to scale). \textit{Left:} Global view where the domain is located at the base of the convective zone (CZ) and at the top of the radiative zone (RZ). \textit{Right:} \tck{Radial representation of} the computational domain 
of extent $\ds d=r_o-r_i \ll r_o$.}%
\label{fig_geo}
\end{minipage}
\end{center}
\end{figure}

In the following we adopt a Cartesian geometry, in a box of \tck{radial} size $\ds d=r_o-r_i$ situated at the base of the convection zone \tck{of a rotating solar-like star (see figure \ref{fig_geo})}. \tck{Our goal is} then to describe locally the evolution of the zonal flow $V$, the toroidal component of the magnetic field $B$ and the potential vector $A$ in this box. A simple 1D description can be obtained by considering the induction equations Eqs.~(\ref{toro}-\ref{polo}) and the Navier-Stokes one Eq.~(\ref{3:eq:EQ_TS_ADIM_1}) averaged in the  azimuthal direction $\phi$ (see Appendix~\ref{3Dto1Dflorentin}). In our case, we do not consider the effects associated with meridional circulation \tck{nor latitudinal fluxes of angular momentum} as the former develop on timescale longer than the magnetic effects \citep{MaederM2003}. \tck{It means that most effects due to rotation are discarded for this first simple approach (see next)}. 
\tck{We acknowledge that the nonlinear interplay between dynamo-generated magnetic fields and meridional circulation could considerably affect the distribution of chemical species and the global evolution of stars \citep{MaederM2005}; this should be the subject of a future paper, beyond the scope of this work.}
In this paper, we limit our attention to dynamical timescales associated with the mean flow and magnetic field evolutions.
\tck{As a consequence of neglecting rotational effects}, the dependency of the fields on the variable $\displaystyle \theta$ has been dropped so as to simplify the study. This can hold if the computational domain lies \tck{far from the poles} or close to the equator which is consistent with the region where we have measured $\ds \alpha_{\phi\phi}$ through the norm in Eq.~(\ref{norme}) \citep[e.g., see][for a 2D approach where the variations with the colatitude are explicitly taken into account]{Jouve2008}. 

\tck{In this framework, the total mean magnetic field $\mathbf{B}$ and velocity field $\mathbf{V}$ read:}
\begin{align}
	\mathbf{B}(r,t) &= \mathbf{\nabla} \times \left(A(r,t)\mathbf{e_\phi}\right) + B(r,t) \mathbf{e_\phi}, \label{eq_B_general}\\
	\mathbf{V}(r,t) &= V (r,t)\mathbf{e_\phi}.
    \label{eq_V_general}
\end{align}
\tck{The governing mean-field equations are thus (see Eqs.~(\ref{toro}-\ref{polo}) and  Appendix~\ref{3Dto1Dflorentin})}:
\begin{align}
\pdt V &= \nu \frac{\partial^2 }{\partial r^2}V -\pdr \left( Rey - Max \right), \label{eqBrunoNS_ok}\\
\pdt A &=  \eta \frac{\partial^2 }{\partial r^2}A + \boldsymbol{\mathcal{E}}\cdot \mathbf{e_\phi},\label{eqBrunoPOLO_ok} \\
\pdt B &= \eta \frac{\partial^2 }{\partial r^2}B+\left(s (\bf{B_P}\cdot\nabla) \frac{\bf{V}}{s}\right)\cdot \mathbf{e_\phi}, \label{eqBrunoTORO_ok} 
\end{align}
where
\begin{itemize}
	\item $\ds \mathbf{B_P}= \nabla \times (A\mathbf{e_\phi})= \frac{\cot \theta A}{r} \mathbf{e_r} - \frac{1}{r}\pdr (rA)\mathbf{e_\theta} $ is the mean poloidal magnetic field. 
	\item $\boldsymbol{\mathcal{E}}\cdot \mathbf{e_\phi} = \alpha_{\phi\phi}B$ is the electromotive force introduced in section \ref{alphaDNS}. Note that \textit{a priori}, two other terms could be considered for $\ds \boldsymbol{\mathcal{E}}\cdot \mathbf{e_\phi}$, \tck{namely} $\alpha_{\phi r} \mathbf{B}\cdot \mathbf{e_r}$ and $\alpha_{\phi\theta} \mathbf{B}\cdot \mathbf{e_\theta}$, as we have access to the whole tensor $\ds \boldsymbol{\alpha}$ from DNS. However, we do not expect those to play a crucial role as in Spruit's theory $B_P\ll B_T$. Assuming that $\ds \alpha_{\phi r}$ and $\ds \alpha_{\phi \theta}$ are comparable (which \tck{is relatively well verified in our SVD measurement from DNS}) we will then discard them but they might be kept in a future study.
	\item $\ds s = r\sin\theta $ is the cylindrical radius which appears due to the spherical geometry.
	\item $\ds Rey$ is the Reynolds stress \tck{representing the radial momentum fluxes carried by IGW}. For a monochromatic wave whose phase velocity is $\ds \omega/k$, \tck{with $\ds \omega$ the wave frequency and $\ds k$ the azimuthal wave number}, this term \tck{can be obtained by using a Wentzel-Kramers-Brillouin (WKB) approximation} \tck{\citep[e.g.,][]{Press81,zahn97}}:
	\begin{equation}
	Rey = \overline{v_r^*v_\phi^*} = F_J \exp\left({-\int_{r_o}^{r}\frac{\gamma}{\left(\omega-kV\right)^4}dr'}\right)- F_J \exp\left({-\int_{r_o}^{r}\frac{\gamma}{\left(\omega+k V\right)^4}dr'}\right),
    \label{eq_Rey_1}
    \end{equation} 
    with $\gamma = k^3\kappa N^3$. \tck{We refer to \citet{Couston2018}'s supplementary material for details of the derivation.} Eq.~(\ref{eq_Rey_1}) describes \tck{the sum of the fluxes of both a prograde and retrograde waves propagating by hypothesis in planes parallel to the equatorial plane (i.e., no latitudinal propagation)}. These waves are assumed to be generated at the base of the convection zone $r_o$ with amplitude $\ds F_J$ and $\ds -F_J$ respectively, and attenuated in the radiative zone over a typical length $\ds \omega^4/\gamma$. Note here that due to the nature of the damping by the radiative regions of stars, this term involves the thermal diffusivity whereas it usually is the molecular viscosity in the atmosphere, where viscous effects dominate the damping. \tck{Eq.~(\ref{eq_Rey_1}) besides considers pure internal gravity waves. Rotation and magnetic field do not affect these waves. Note that because we focus on the interplay between Tayler-Spruit dynamos and the action of the waves, $\ds Rey$ only contains the latter, but as described in \citet{DanielPG23}, it could also include some turbulence as the one induced by the former.}
	\item $\ds Max$ is the Maxwell stress resulting from the magnetic field retroaction on the velocity. We assume that it is of the form $\ds Max = \cot \theta \frac{AB}{\rho_0 \mu_0 r}$, $\ds \rho_0$ being the fluid density which is assumed constant.

\end{itemize}

The fact that $\ds \theta$ remains here comes from the local expression of the MHD equations in spherical geometry. Varying the latitude here would mean that we rotate the "box" which would result in changing the prefactor of the Maxwell stress for instance. The range of latitude reachable must therefore stay near the equator as we would expect huge variations of $\ds V$ notably close to the poles, \tck{which would go against the assumptions of no-dependence in the $\ds \theta$ direction}. We cannot however situate the domain exactly at the equator as some term would cancel due to the prefactor $\ds \cot \theta$. \tck{Following the discussion in section \ref{section_discussion_TS}, it means that to describe the equatorial dynamics we should account for some meridional circulation, which is beyond the scope of the present work.} In practice, we did not vary $\ds \theta$ and we worked for all the work reported here at $\ds \theta = \pi/3$. \tck{Note that this is out of the range of where we measured $\mathcal{N}(\alpha_{\phi\phi})$ (Eq.~(\ref{norme})) as the point of the present work is to extend global DNS results to a local approach.}

Introducing $l_0=d$, $t_0 = d^2/\kappa$, $B_0=\sqrt{\rho\mu}\kappa/d$ respectively the length, time and magnetic field scales, the previous equations Eqs.~(\ref{eqBrunoNS_ok},\ref{eqBrunoPOLO_ok} and \ref{eqBrunoTORO_ok}) are made dimensionless as:
\begin{align}
\frac{\partial \widetilde{V}}{\partial \widetilde {t}}  &= Pr \frac{\partial^2 }{\partial \widetilde{r}^2}\widetilde{V} -\frac{\partial}{\partial \widetilde{r}} \left(F \widetilde{Rey} - \frac{\cot \theta \widetilde{A} \widetilde{B}}{\widetilde{r}} \right), \label{eqBrunoNS_okadim}\\
\frac{\partial \widetilde{A}}{\partial \widetilde {t}} &=  \frac{Pr}{Pm} \frac{\partial^2 }{\partial \widetilde{r}^2}\widetilde{A} + \cot \theta \widetilde{\alpha(\widetilde{r},\widetilde{t})} \widetilde{B},\label{eqBrunoPOLO_okadim} \\
\frac{\partial \widetilde{B}}{\partial \widetilde {t}} &= \frac{Pr}{Pm} \frac{\partial^2 }{\partial \widetilde{r}^2}\widetilde{B}+\cot \theta \widetilde{A} \frac{\partial}{\partial \widetilde{r}}\frac{\widetilde{V}}{\widetilde{r}},\label{eqBrunoTORO_okadim} 
\end{align}
with: 
\begin{equation}
	\widetilde{Rey} =  \exp\left({-\int_{r_o/d}^{r/d}\frac{L^{-1}}{\left(1-D \widetilde{V}\right)^4}dr'}\right)- \exp\left({-\int_{r_o/d}^{r/d}\frac{L^{-1}}{\left(1+D \widetilde{V}\right)^4}dr'}\right),
    \label{eq_rey_adim}
\end{equation}

\noindent \tck{and where dimensionless quantities have been noted as $\ds \widetilde{..}$.}
We have introduced the following additional parameters:
\begin{itemize}
    \item $\displaystyle L = \frac{\omega^4}{d\gamma }=\frac{\omega^4}{d k^3\kappa N^3}$ quantifies the typical length over which the waves will be damped (compared to the size of the domain).
    \item $\displaystyle D=\frac{\kappa/d}{\omega/k}$ is the ratio between the \tck{velocity scale} of the flow ($\ds \kappa/d$) and the phase velocity of the waves ($\ds \omega/k$).
    \item $\displaystyle  F=\frac{F_J}{\kappa^2/d^2}$ can be seen as the ratio between the \tck{momentum of the waves at the base of the convective zone in $\ds r=r_o$---the prefactor $\ds F_J$ of the Reynolds tensor---to the large scale characteristic kinetic energy of the flow---$\ds (\kappa/d)^2$ with our choice of velocity scale.} \tck{It is a measure of the two competing physical phenomena of the problem, the forcing term coming from the waves and dissipation due to diffusion}.
    \item $\ds Pr=\frac{\nu}{\kappa}$ and $\ds Pm=\frac{\nu}{\eta}$ are the \tck{thermal and magnetic Prandtl numbers, respectively.}
\end{itemize}

The domain of integration is bounded by $r_o$ the size of the radiative zone on top and by $r_i$ below (figure \ref{fig_geo}). The choice of $r_i$ is non-trivial, as it introduces a new length scale to the problem $\ds d=r_o-r_i$. For this model, we want $d \ll r_o$ so that it corresponds to a local box confined at the top of the radiative zone, but $\ds d$ must also be large compared to the length over which the waves will be damped $\ds \omega^4/\gamma$ so that the oscillation can be maintained. In this spirit, we work at $\ds r_o/d = 10$ \tck{and $L=0.25<1$}. \tck{The chosen value of $\ds L$ is consistent with previous works \citep[e.g.,][]{RenaudNadeau2019}; the size of the box is 4 times the length over which the waves are damped.} \\

The boundary conditions we consider are no slip with $\ds V=0$ for the velocity at $\ds\widetilde{r}=r_o/d$ as we set ourselves in the reference frame of \tck{the boundary between the convective and radiative zones}, the so-called tachocline, and stress-free at $\ds \widetilde{r}=r_i/d$, enforcing $\ds \partial V /\partial r =0$. The latter is motivated by the fact that we want the flow to evolve freely at the bottom of the radiative zone. For the magnetic field, as we want to focus on the possibility of a field to develop in the radiative zone, we impose $\ds B=A=0$ at $\ds \widetilde{r}=r_o/d$ so as not to consider any magnetic influence to or from the convective zone. At $\ds \widetilde{r}=r_i/d$, we also enforce $\ds B=A=0$ \tck{as we expect the magnetic field to be localised where the flow gradients are strong, i.e. close to $\ds r_o$}. 

The last piece of information missing is the choice of the function $\ds \widetilde{\alpha(\widetilde{r},\widetilde{t})}$, which is connected to the results obtained in DNS (see section \ref{alphaDNS}). We write it as:

\begin{equation}
	\widetilde{\alpha (\widetilde{r},\widetilde{t})} = f(\widetilde{r},\widetilde{t})\widetilde{\alpha_{1D}}.
	\label{eq_forme_alpha}
\end{equation}

The first term in the right hand side (RHS) of Eq.~(\ref{eq_forme_alpha}) is a function $\ds f$ which accounts for the activation of the $\ds \alpha-$effect. Indeed, this term cannot always be present in the equations or it would mean that a dynamo is possible as long as a mean flow develops. Motivated by the conditions for Tayler instability to set in, Eqs.~(\ref{eq:criteres A}) and (\ref{eq:criteres TI}), we impose that $\ds f$ becomes non zero when \tck{the following} conditions are respected:

\begin{align}
	&\frac{d \log \widetilde{B}}{d \log \widetilde{s}}>1, \label{condition1}\\
	&\frac{\widetilde{B}}{\sqrt{\widetilde{r}}} > \left ( \tau_\kappa \omega\right)^{1/6}\left(\frac{d}{r_o}\right)^{1/2}\left(\frac{Pr}{Ek}\right)^{1/4}\left(\frac{Pr}{Pm}\right)^{1/2}\frac{1}{D^{1/2}L^{1/6}}, \label{condition2}
\end{align}

\begin{figure}[h]
\begin{center}
\begin{minipage}{120mm}
{\resizebox*{12cm}{!}{\includegraphics{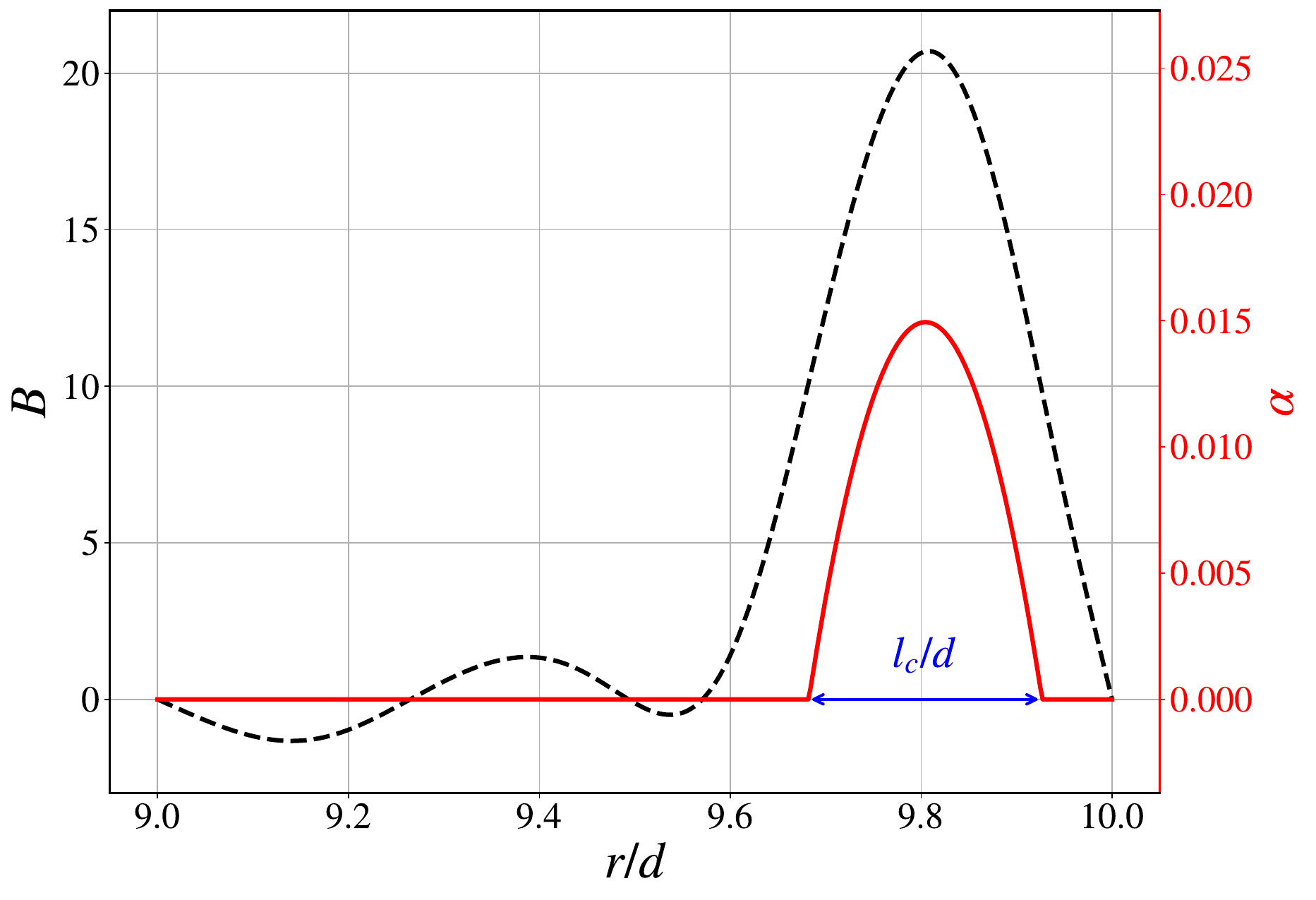}}}%
\caption{Typical profile for $\ds B$ (left, dashed black) and $\alpha$ (right, red) (the values are in code units). The waves are excited on the right of the domain, at $\ds r/d=10$.}%
\label{fig_profile_alpha}
\end{minipage}
\end{center}
\end{figure}

\noindent \tck{with $\ds \tau_{\kappa}:= d^2/\kappa$ a thermal diffusion time}. \tck{Concretely, we converted the criteria of Eqs.~(\ref{eq:criteres A}) and (\ref{eq:criteres TI}) which are expressed with dimensions into the dimensionless system used in the present 1D model.} Besides these first criteria (Eqs.~(\ref{condition1}-\ref{condition2})), we \tck{proceed similarly to} enforce a spatial dependency. For the radii where conditions (\ref{condition1}) and (\ref{condition2}) are verified, we impose that $\ds \widetilde{\alpha}$ be set equal to an order 2 polynomial whose roots will be enforced to be separated by Tayler's critical length (Eq.~\ref{lc_spruit}) and centred on such a radius (see figure \ref{fig_profile_alpha}):

\begin{equation}
	\frac{l_c}{d} = \left ( \tau_\kappa \omega\right)^{-1/6}\left(\frac{Pr}{Ek}\right)^{1/4}D^{1/2}L^{1/6}.
	\label{lc code}
\end{equation}
The second term $\ds \widetilde{\alpha_{1D}}$ in the RHS of Eq.~(\ref{eq_forme_alpha}) is the amplitude of this parameterised $\ds \alpha-$effect. It is controlled by the dependency obtained with global simulations. \tck{Noting that it has the dimension of a velocity, we have to transform the dimensionless system of 3D DNS to the one of the present 1D model:}
 
\begin{equation}	
	\widetilde{\alpha_{1D}} = \frac{V_{0,DNS}}{V_{0,1D}}\widetilde{\alpha_{DNS}},
\end{equation} 

\noindent \tck{with $\ds V_{0,DNS} = r_i \Delta \Omega$, $\ds V_{0,1D} = \kappa/d$ and $\ds \widetilde{\alpha_{DNS}} = \mathcal{N}(\alpha_{\phi\phi})$. Eventually we get:}
 
\begin{equation}
	\widetilde{\alpha_{1D}} = \beta Ek^a Pm^d \left( \frac{(\tau_\kappa \omega)^{2/3}}{Pr}\frac{1}{D^2 L^{2/3}} \right) ^b \frac{1}{Pr^c}\left(\max_{r} \widetilde{\Omega} - \min_{r} \widetilde{\Omega}\right)^{1+c},
	\label{alpha 1D}
\end{equation}

\noindent where $\ds \beta$ is a prefactor which depends only on geometric factors \tck{that are due to different length units in the DNS and 1D model. It is} equal to $\ds 5$. \tck{We note that the first term in brackets in the RHS of the last equation is the Rayleigh number expressed as a function of the other parameters of the problem.} The major difference between the 1D model and the 3D DNS is how the differential rotation is handled. In DNS, we impose a large scale shear through the mean of the velocity boundary conditions---it is a Couette problem. Here, the source of differential rotation is directly given by the transport of angular momentum by waves generated at the top boundary, leading to the shear-layer oscillation. Therefore, the $\ds \Delta \Omega$ of the DNS is set equal in the 1D code to the last term in Eq.~(\ref{alpha 1D}), $\ds \max_{r} \widetilde{\Omega} - \min_{r} \widetilde{\Omega}$, which is now a function of time, with \tck{$\ds \widetilde{\Omega} = \widetilde{V}/\widetilde{s}$}. \tck{The prefactor $\ds \cot \theta$ in front of the $\ds \alpha$ term in Eq.~(\ref{eqBrunoPOLO_okadim}) is to assure that our equations respects the equatorial symmetry.}\\

A comment need be made on the absence of quenching on the amplitude in Eq.~(\ref{alpha 1D}). Indeed, classical models \citep{moffatt,Cattaneo1996a} suggest that once the mean toroidal field becomes "important enough", the $\ds \alpha-$term should be quenched, i.e. reduced. This is to describe the fact that such an important magnetic field should be associated with a braking of the flow and a decrease of the kinetic helicity to which classical models relate $\ds \alpha$. This usually takes the form of $\ds \alpha \propto \alpha_0/(1+\mu B^2)$ with $\ds \alpha_0$ its typical amplitude and $\ds \mu$ a constant whose value is still debated \citep{Proctor2003}. The key point is that this quenching coming from a kinematic study, it is not needed here as the Lorentz force is included to act on the flow, and thus reduces the effective differential rotation and eventually $\ds \widetilde{\alpha_{1D}}$ (Eq.~\ref{alpha 1D}). 

In describing the $\ds \alpha$-effect implemented in our model, two other parameters have been introduced. The first one is $\ds \tau_\kappa \omega = \omega d^2/\kappa$, the ratio of the thermal diffusion time to the typical timescale of the waves $\ds \omega^{-1}$, which is expected to be larger than unity in the radiative zone of the Sun. In all the simulations reported here it will be set to $\ds 10^3$. 

The second one is the Ekman number $\ds Ek$ which controls how the Tayler-Spruit dynamo will develop through the value of the threshold on the amplitude of $\ds B$ (Eq.~\ref{condition2}), the Tayler critical lengthscale (Eq.~\ref{lc code}) and the amplitude of the $\ds \alpha$ term (Eq.~\ref{alpha 1D}). It therefore introduces an effective rotation in the problem. This is yet another hypothesis of the problem where we assumed that the meridional circulation which would account for the Coriolis force in Eq.~(\ref{eqBrunoNS_ok}) is weak compared to the zonal flow we focus on \citep{MaederM2003}. In the following simulations, its value will be set to $10^{-5}$. 

\tck{We can from these last two parameters define the ratio of the Brunt-Väisälä $\ds N$ to the overall rotation $\ds \Omega_o$ as $\ds N/\Omega_o= (\tau_\kappa \omega/L)^{1/3} Ek (r_o/d)^2/(DPr)$. A list of all the parameters is presented in Table \ref{tab_param} with some typical values for our 1D model as well as more realistic values corresponding to the top of the solar radiative zone.}

\section{Towards stellar applications}
\label{section_results}

We now integrate Eqs.~(\ref{eqBrunoNS_okadim}-\ref{eqBrunoTORO_okadim}) \tck{using a numerical model inspired from the one described in \citet{Renaud2018PHD}, employing an explicit scheme for the time evolution and a centred one for the spatial derivatives}.

\tck{In the following, we fix $\ds Pr=0.01$, $\ds L=0.25$ and $\ds D=3.10^{-3}$ and vary $\ds F$. \tck{As discussed in the previous section, we also recall that we have set $\ds r_o/d = 10$, $\tau_\kappa \omega = 10^3$ and $Ek=10^{-5}$.} Although these values are far from values expected in a stellar environment such as the top of the solar RZ (see Table \ref{tab_param}), they enable us to investigate the effect of coupling the SLO with a dynamo at a small numerical cost. Indeed, the large values of $Pr$ and $Ek$ affect the Tayler length scale (Eq.~\ref{lc code}), so the values of $\nu$ and $\kappa$ used in their definition can be seen as effective turbulent diffusion coefficients. Despite this limitation which will be focused on in a future work, we emphasise that we integrate the present 1D model with a value of $\ds N/\Omega_o=529$. The simulated 1D environment is therefore much more stratified than what was reached in DNS ($O(1-10)$), besides being close to the value expected for the top of the Sun's radiative zone ($O(10^3)$).}

\tck{For the sake of simplicity, we drop the $\ds \widetilde{..}$ over the dimensionless variables of the problem in what follows.} We describe the results obtained \tck{by defining key output variables.}

\begin{table}[t]
\centering
\begin{tabular}{l|l|l|l}
                       & Expression & 1D model & top of the Sun's RZ \\
                       \hline
Independent parameters & $\ds F = \frac{F_J}{\kappa^2/d^2}$  & $\ds [10^2-10^3]$  & $\ds 10^{12}$\\
                       & $\ds D = \frac{\kappa/d}{\omega/k}$ & $3.10^{-3}$  & $\ds 10^{-8}$\\
                       & $\ds L = \frac{\omega^4}{k^3 \kappa N^3 d}$ & $\ds 0.25$   & $\ds 10^{-1}$\\
                       & $\ds Pr=\nu/\kappa$ & $\ds 10^{-2}$  & $\ds 10^{-6}$\\
                       & $\ds Pm=\nu/\eta$   & $\ds 10$  & $\ds 10^{-2}$ \\
                       & $\ds \tau_\kappa \omega = (d^2 \omega)/\kappa$ &  $\ds 10^3$    & $\ds 10^7$\\
                       & $\ds Ek=\frac{\nu}{r_o^2\Omega_o}$ & $\ds 10^{-5}$  & $\ds 10^{-15}$\\
                       \hline
Relation with other parameters & $\ds \frac{\omega}{N}=L^{1/3}D\left(\tau_\kappa \omega\right)^{2/3} $ &   $\ds 0.2$     & $\ds 10^{-3}$       \\
                       & $\ds \frac{N}{\Omega_o}= (\frac{\tau_\kappa \omega}{L})^{1/3} Ek \frac{(r_o/d)^2}{DPr} $ &  $\ds 529$   & $\ds 10^3$\\
\end{tabular}
\caption{List and values of the dimensionless parameters used in the 1D model \tck{($r_o/d=10$)}. Estimated values are also given at the top of the radiative zone of the Sun, following numerical values reported in \citet{Press81, KimM01, Garcia2007, Rogers2010, Schumacher2020}.}
\label{tab_param}
\end{table}

\tck{Besides the classical energies averaged over the volume}:
\begin{equation}
	E_X := \frac{1}{2} \int_{r_i/d}^{r_o/d} X(r)^2dr,
\end{equation}
$X$ being either $\ds V, A$ or $\ds B$, we also introduce the following probes in the bulk of the domain:
\begin{equation}
	X_1 := X(r_o/d-L/2,t) \quad \text{and} \quad X_2 := X(r_o/d-3L,t).
\end{equation}

\tck{Changing the positions of these probes does not affect the interpretation we make in what follows.}

\begin{figure}[t]
\begin{center}
\begin{minipage}{150mm}
{\resizebox*{15cm}{!}{\includegraphics{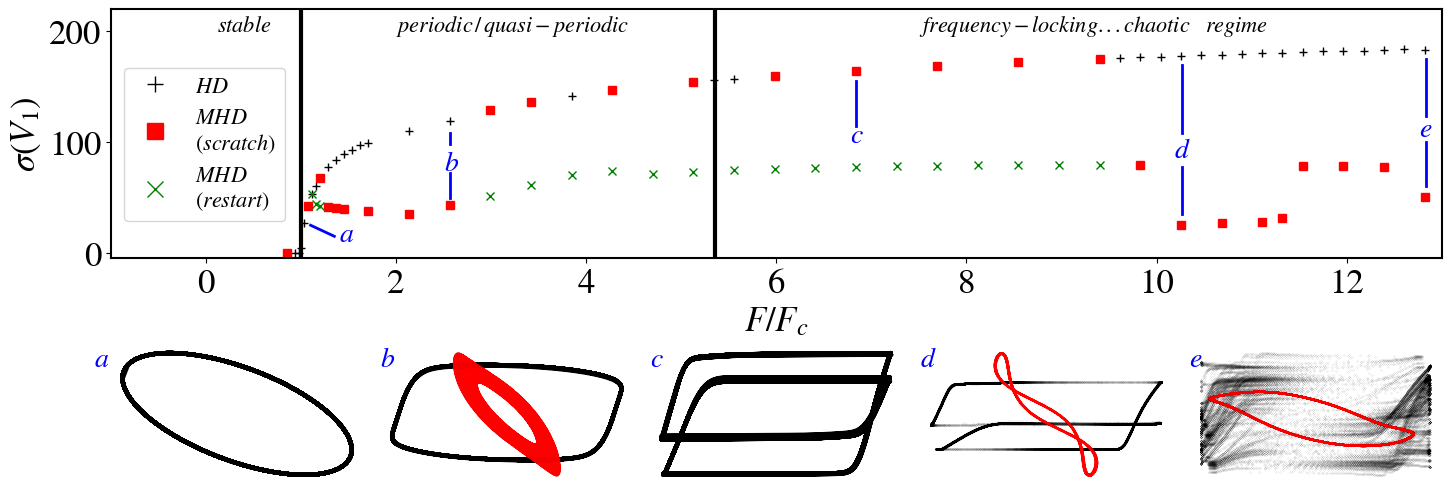}}}%
\caption{Bifurcation diagram of $\ds \sigma(V_1)$ the standard deviation of $V_1$ (see text for definition) for $Pr=0.01,Pm=10,Ek=10^{-5},L=0.25, D=3.10^{-3}$ and increasing $\ds F$, normalised by the measured critical value of the onset of oscillations $\ds F_c=117$. \textit{a-e}: Phase portraits in the $\ds (V_1,V_2)$ plane for $\ds F/F_c=\left\{1,2.56,6.84,10.26,12.82\right\}$ (a,b,c,d and e respectively). \tck{Simulations are here reported without (HD, black crosses) or with magnetic field, being either initialised from the condition described in the text (MHD scratch, red squares) or restarted from a previously computed solution (MHD restart, green crosses).}}%
\label{fig_regime}
\end{minipage}
\end{center}
\end{figure}

Figure \ref{fig_regime} shows an example of bifurcation diagram for the standard deviation of $\ds V_1$, \tck{$\ds \sigma(V_1):= \sqrt{\left<\left( V_1-\left<V_1\right>_t \right)^2\right>_t}$} as $\ds F$ varies, every other parameter being fixed, with or without including magnetic effects. $\ds \left<..\right>_t=1/\tau_\kappa\int_{t_{statio}}^{t_{statio}+\tau_\kappa}..dt$ stands for a time average once a stationary state $t_{statio}$ is reached and is measured on one thermal diffusive time $\tau_\kappa$. The bifurcation diagram illustrates the wide variety of dynamical regimes. \tck{In the hydrodynamical regime, the mean flow oscillation was notably observed by \citet{Plumb1978} in a laboratory experiment and later characterised numerically by \citet{Yoden1988} as a Hopf bifurcation. Following recent developments for a laboratory experiment by \citet{semin16} in the context of the Quasi-Biennial Oscillation (QBO), a weakly nonlinear computation was proposed by \citet{semin2018} in an infinite domain and showed that depending on \tck{the relative influence of diffusion and linear damping}, this Hopf bifurcation could either be sub or supercritical, the threshold being a function of the other parameters $\ds L,D$ and $\ds Pr$. In our case, as we do not consider linear damping on the momentum equation, we do not expect any subcriticality, at least in the non-magnetised case.}

\subsection{Dynamics of the wave-induced oscillation}
\label{section_results_MHD}

Before describing the novelty of our approach by highlighting magnetic influences, \tck{we will first} recall purely hydrodynamical ($\ds A=B=0$) results  illustrated in figure \ref{fig_regime} in black. The initial velocity is taken linear \tck{as a function of the radius} and arbitrarily weak ($5.10^{-2}$ in code units at $r=r_i/d$ and $0$ at $r=r_o/d$). We do not observe any dependency on the shape of the initial profile for the saturated solutions. 
In the case of the sum of a prograde and a retrograde monochromatic wave with symmetric forcing, i.e. corresponding to $\ds F$ and $\ds -F$ prefactor respectively, $\ds V=0$ is a stable solution of Eq.~(\ref{eqBrunoNS_okadim}) for moderate values of $\ds F$ (see figure~\ref{fig_regime} black crosses). This can be understood qualitatively by saying that in a stationary state, the diffusion term is stronger than the Reynolds stresses and prevents a non-zero solution from existing. However, if $\ds F$ is increased above a threshold $\ds F_c$, \tck{this balance does not hold anymore and} a non-zero mean flow can be driven by the flux of the two waves which will lead to an oscillating solution. The inset ($\ds a$) of figure \ref{fig_regime} exhibits the limit cycle of the phase portrait of $\ds V_1$ and $V_2$ close to this onset. 
Increasing the control parameter $\ds F$ past the first onset leads to a modification of the limit cycle of figure~\ref{fig_regime}($\ds a$), with the emergence of harmonics in the Fourier spectrum of a given probe (figure~\ref{fig_regime}($\ds b$) in black). The limit cycle will afterwards undergo a transition leading to the locking of the dominated period to new ones (figure~\ref{fig_regime}($\ds c$)), through several phase-locking episodes (figure~\ref{fig_regime}($\ds d$) in black for the emergence of the quarter of the dominant frequency) before eventually transitioning to a chaotic regime (figure \ref{fig_regime}($\ds e$)). The emergence of peculiar frequencies on the way to chaotic regime was first observed by \citet{KimM01}. The successive frequency locking events leading to chaos were elegantly displayed by \tck{\citet{RenaudNadeau2019}}. 
The mechanism of the SLO therefore hosts a wide range of hydrodynamical regimes, either periodic or quasi-periodic, phase-locking events or even chaos. Adding magnetic field will modify this picture.

\subsection{Coupling with the dynamo}

Having identified the regions of the parameter space corresponding to different flow regimes, we now add a non-zero magnetic field initially, whose profile is once again taken linear with arbitrary weak amplitudes ($\ds 10^{-1}$ in code units for $\ds A$ and $\ds B$).
\tck{The result is represented by} red squares in figure \ref{fig_regime}. Below $\ds F_c$ we unsurprisingly obtain a null solution as the differential rotation needed to "activate" the magnetic instability is not present. However, simply reaching $\ds F_c$ is not sufficient to obtain a dynamo. Indeed, the first magnetic solution we obtain is for $\ds F/F_c = 1.154$, which we attribute to the fact that the differential rotation of the mean flow is not sufficient before to trigger and maintain the \tck{parameterised} dynamo. Beyond this threshold, several oscillating solutions exist (see insets $b,d$ and $e$ of figure \ref{fig_regime}).

\begin{figure}[t]
\begin{center}
\begin{minipage}{160mm}
{\resizebox*{16cm}{!}{\includegraphics{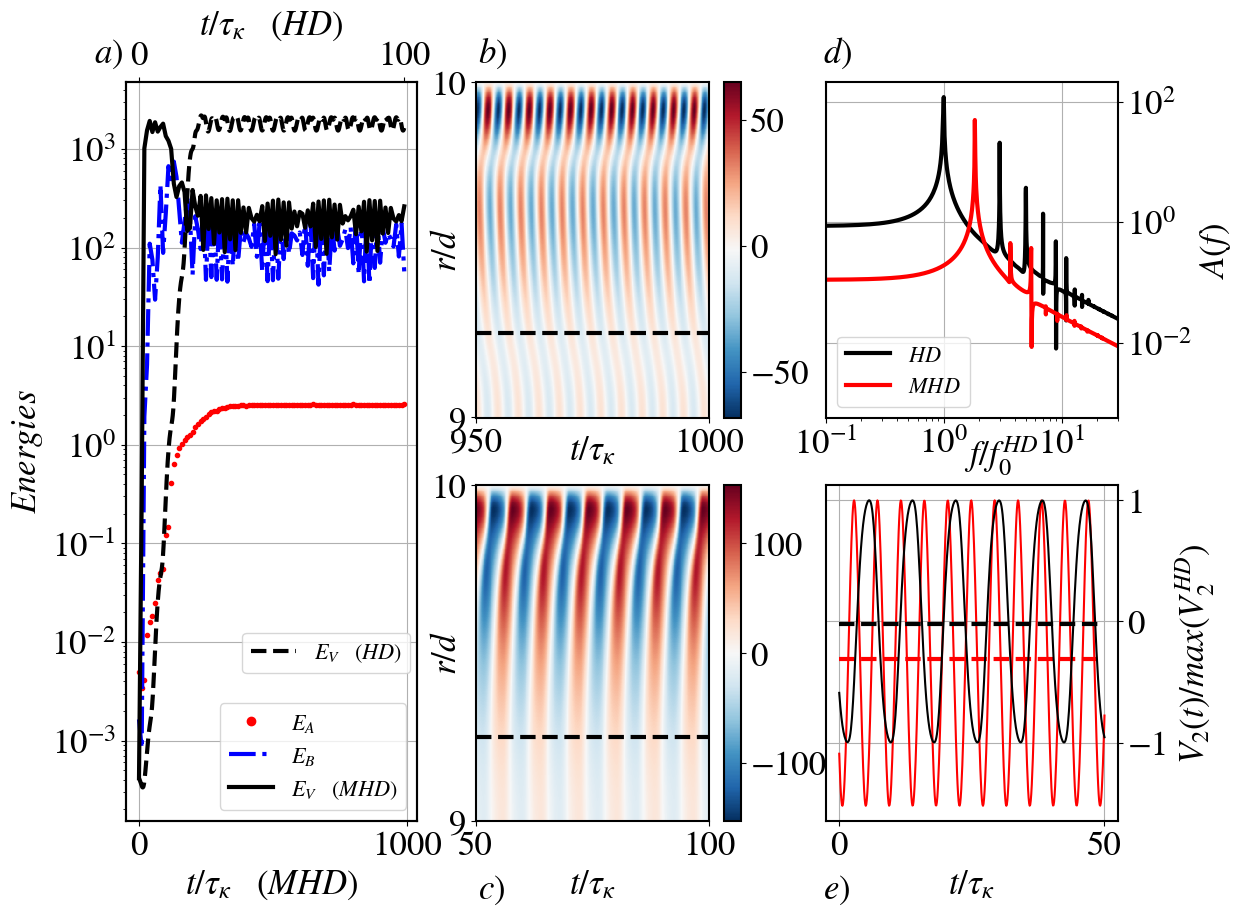}}}%
\caption{Comparison of the simulation at $Pr=0.01,Pm=10,Ek=10^{-5},L=0.25,D=3.10^{-3},F=200$ with (MHD) or without magnetic field (HD). \textit{a:} Timeseries of the energies. \textit{b and c:} Hovmöller diagrams of the velocity field for the MHD case (b) and the HD (c) one. The dashed line corresponds to the location of the second probe $\ds X_2$. \textit{d:} Fourier spectra of $\ds V_1$. \textit{e:} Time evolution of $\ds V_2$ in the two cases.}%
\label{figure time series}
\end{minipage}
\end{center}
\end{figure}

Figure \ref{figure time series}($\ds a$) shows typical time series for $\ds F/F_c = 1.71$ both with or without magnetic field. For the former, as the mean flow quickly sets in ($\ds t/\tau_\kappa <100$), it leads to the generation of $\Omega-$effect and thus to the amplification of $\ds B$ (Eq.~\ref{eqBrunoTORO_okadim}). Once the threshold for Tayler instability is reached by $\ds B$, the parameterised $\ds \alpha$-effect becomes non-zero, and in turn amplifies $\ds A$, therefore again amplifying $\ds B$. This is a classical $\ds \alpha-\Omega$ dynamo loop. The natural quenching here appears once $\ds B$ becomes strong enough ($\ds t/\tau_\kappa \approx 100$). Indeed, at this point, the Maxwell stress is sufficient to act back on the mean flow (Eq.~\ref{eqBrunoNS_okadim}) which leads to a decrease of the overall kinetic \textit{and} magnetic toroidal energy. This is due to the fact that as the flow is slowed down by the Lorentz force, the $\Omega-$effect is reduced and gives rise to saturation of the dynamo. \tck{Such a retroaction of the flow on the magnetic field was also observed in the low-dimensional model proposed in \citet{DanielPG23} but with a different outcome as it indeed gave rise to an amplification of the turbulent velocity field. This difference is notably due to the fact that the Reynolds stress term used in the present study only models the action of the waves and not the turbulence induced by the dynamo: in the present case, $V$ represents the mean flow. Besides being a source of turbulence, dynamo action can lead to subcritical behaviours, which are not \textit{a priori} expected here.}

Focusing on the observed magnetic oscillating solutions reported here, their stability region is however not trivial as for $\ds F/F_c \in \left [ 3,9 \right ] $ we did not obtain a dynamo starting from the initial condition described before \tck{nor by varying the amplitude of the initial field}. Non-zero magnetised solutions could be obtained but by restarting a simulation from a previously saturated computation at a different $\ds F/F_c$ (see the green crosses in figure \ref{fig_regime}) which led to the emergence of several branches. Interestingly, for $\ds F/F_c = 12.82$ (figure \ref{fig_regime}($\ds e$)) we observe an oscillating solution in a region where the purely hydrodynamical solution is chaotic. Adding magnetic field therefore changes the successive onsets of different hydrodynamical solutions described earlier. The magnetic field can even in some cases laminarise the flow. \tck{This is comparable with previous results from \citet{KimM03} who, in contrast to our model, studied the effect of a uniform toroidal magnetic field directly on the internal gravity waves (through the dispersion relation) and observed that the mean shear profile would be smoother in the presence of the magnetic field.} \citet{Cattaneo1996b} also observed such an effect in a different context when describing the differences between kinematic and dynamic dynamos; if the flow of the former exhibited chaotic behaviour, taking into account the back reaction of the field on the flow for the latter led to a drastic suppression of its chaotic properties; \tck{ the retroaction of the Lorentz force decreases the amplitude of the flow, leading the magnetised system to evolve with a reduced effective velocity amplitude. As chaotic regimes are obtained in the hydrodynamical limit when the velocity of the flow is important enough, this qualitative description suggests that in the magnetised case the amplitude of the velocity is too small to reach chaotic attractors. Note that magnetised chaotic solutions could be obtained but at a higher value of the control parameter $F$, the presence of the magnetic field simply shifting the values of $F$ at which chaos is reached.} \tck{Our results are however different from the work of \citet{Rogers2010} who showed by imposing a toroidal magnetic field on the stably stratified layer that the resulting flow could become time independent; here all the magnetised solutions are oscillatory.}

\begin{figure}[h]
\begin{center}
\begin{minipage}{100mm}
{\resizebox*{10cm}{!}{\includegraphics{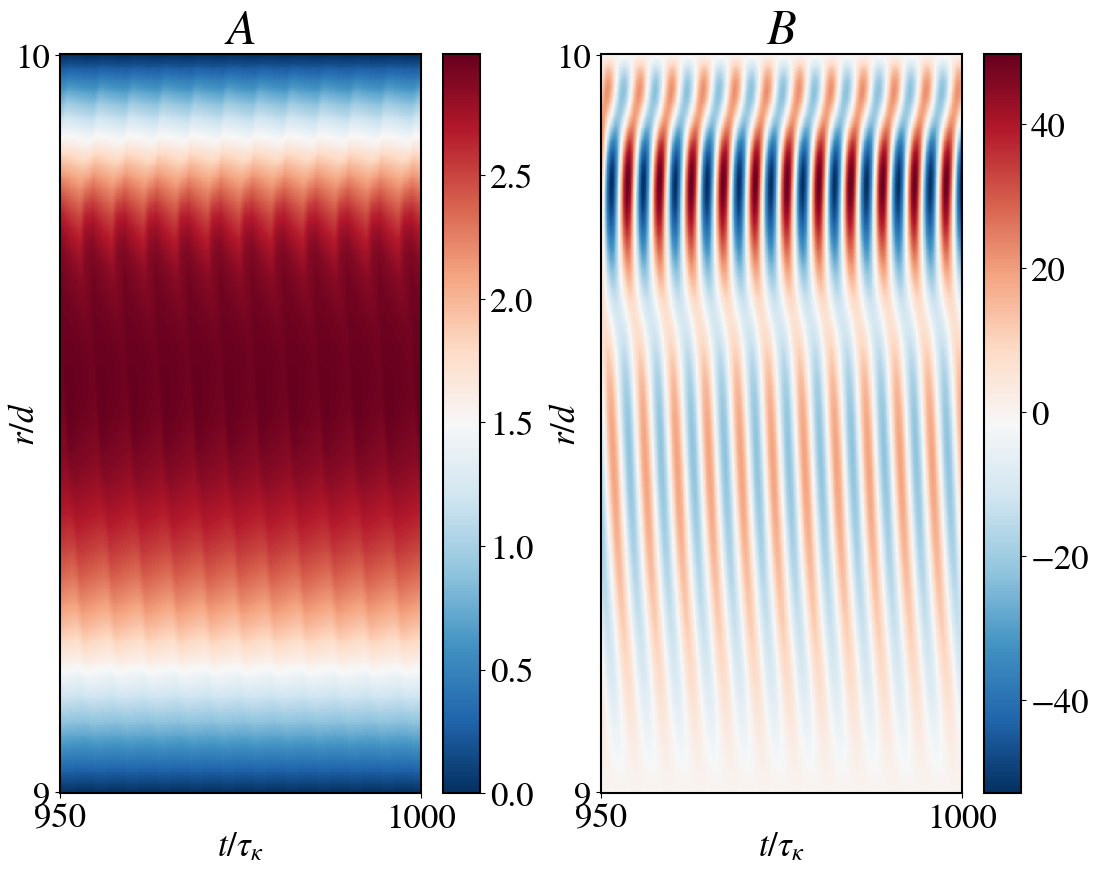}}}%
\caption{To complement figure \ref{figure time series}, we also report Hovmöller diagrams of $\ds A$ and $\ds B$ for the simulation at $Pr=0.01,Pm=10,Ek=10^{-5},L=0.25,D=3.10^{-3},F=200$.}%
\label{figure time series AB}
\end{minipage}
\end{center}
\end{figure}

\tck{Figures \ref{figure time series}($\ds b$ and $\ds c$) compare the Hovmöller---space and time---diagrams of $\ds V$ with and without magnetic fields, respectively, for the same simulation parameters. Although the MHD solution is still periodic, the associated period is nearly divided by two compared to the hydrodynamical solution. This is confirmed by the Fourier spectra of $\ds V_1$ computed once a stationary state has been reached (see figure \ref{figure time series}($\ds d$)).  Now, focusing on the flow amplitude, figure \ref{figure time series}($\ds e$) displays timeseries of $\ds V_2$ normalised by its maximum in the HD case for the two cases. In the hydrodynamical regime, the mean goes to $\ds 0$ (black dashed line) but does not when the dynamo sets in. We attribute this to the fact that in the former case, the equations are invariant with respect to the symmetry $\ds V \rightarrow -V$, but this symmetry is broken in the latter. This leads the mean flux of stresses to being non-zero locally so as to deposit angular momentum and strongly affects the evolution of the flow. The figure also illustrates the drastic reduction of the amplitude of the flow in the MHD case. Such a decrease merely results from the quenching induced by the additional Maxwell stress. To complement figure \ref{figure time series}, we also show the diagrams for $\ds A$ and $\ds B$ in the magnetic case in figure \ref{figure time series AB}, for the same simulation parameters as in figure \ref{figure time series}}.

\begin{figure}[h]
\begin{center}
\begin{minipage}{150mm} 
\subfigure[]{
\resizebox*{7.5cm}{!}{\includegraphics{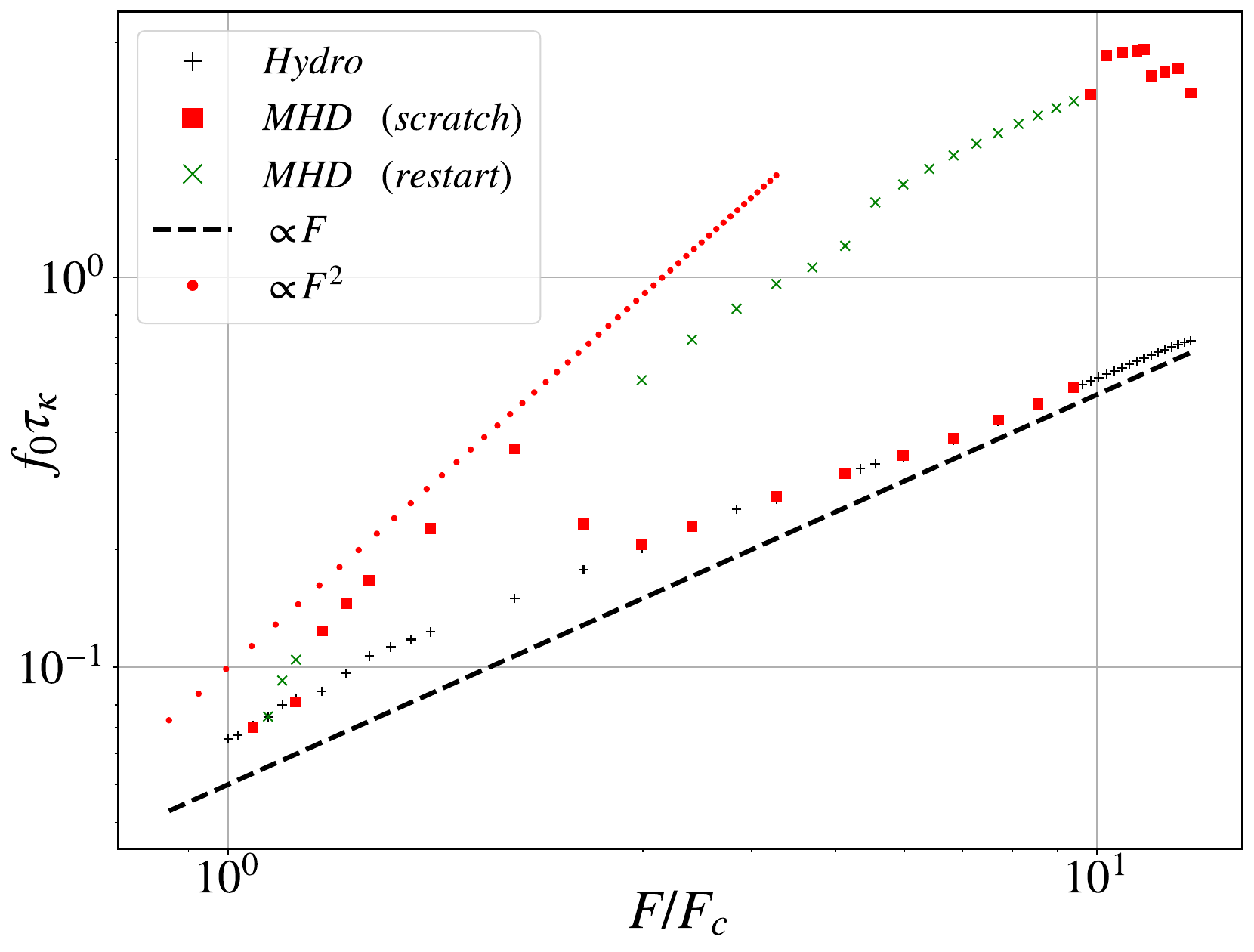}}}%
\subfigure[]{
\resizebox*{7.5cm}{!}{\includegraphics{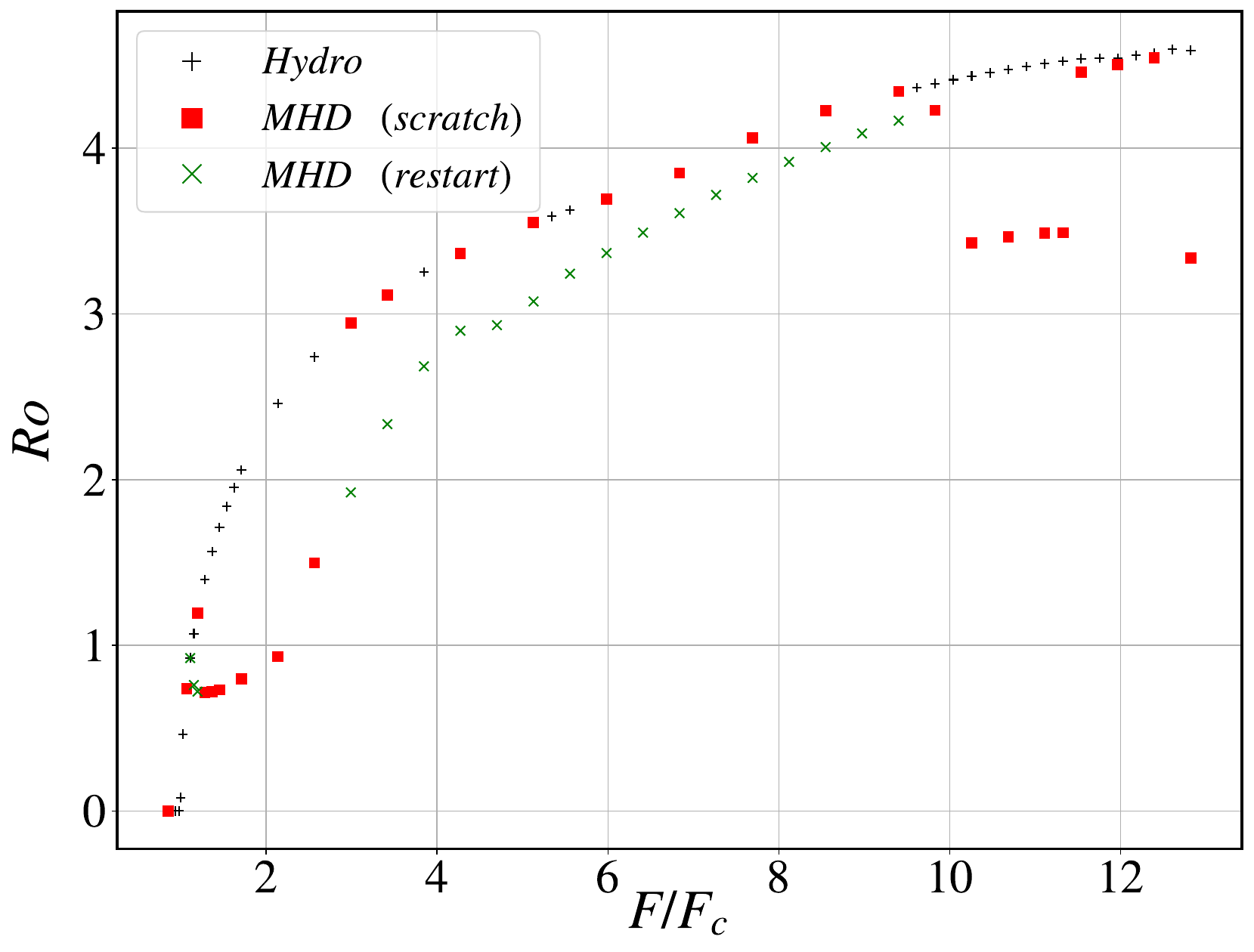}}}%
\caption{Quantitative differences between the MHD and HD solutions of the runs reported in figure \ref{fig_regime}. \textit{a:} Dominant frequency measured for each run by Fourier transform. The two branches exhibit a clear distinct behaviour. \textit{b: }Rossby number $\ds Ro = \delta V/d \Omega_o$ measured for the same simulations, again displaying a different behaviour whether magnetic field is present or not. In the former, $\ds Ro$ is reduced compared to the latter.}%
\label{fig_bifu_f_Ro}
\end{minipage}
\end{center}
\end{figure}

\tck{Figure \ref{fig_bifu_f_Ro}($\ds a$) quantifies the change of period by showing the bifurcation of the dominant frequency for each simulation. We see that the dominant frequency increases in general as $\ds F$ increases on both the HD and MHD branches. Indeed, with $ \ds F$ becoming more important, the transfer of momentum from the IGW to the mean flow gets faster, resulting to an increase of the velocity oscillation. Moreover, we also see that the magnetic field usually amplifies the frequency, which we will discuss later.}

Adding magnetic field \tck{therefore} strongly affects the evolution of the flow and the transport of angular momentum. Besides the behaviour observed in figure \ref{figure time series}($\ds e$), the changes concerning the latter can be quantified in terms of a Rossby number $\ds Ro := \delta V/d \Omega_o$ whose bifurcation is reported in figure \ref{fig_bifu_f_Ro}($\ds b$). $\ds \delta V$ stands here for the time average value of $\ds \max_{r} V - \min_{r} V$. The MHD solution exhibits a reduced value of $\ds Ro$ compared to the hydrodynamical one. Even if the discrepancy between the two seems to diminish as $\ds F$ increases, a highly supercritical branch does seem to maintain a reduced value of $\ds Ro$. 
\tck{The fact that $Ro$ is reduced with the magnetic field can give some explanations as to why the oscillation in the MHD case is increased compared to the purely hydrodynamical one. Indeed, the form of the Reynolds stresses assumed in Eq.~(\ref{eq_rey_adim}) describes the action of two waves, a prograde one propagating in the sense of the zonal flow and a retrograde one propagating in the opposite sense. The damping rate as a function of the travelled distance is proportional to $\ds (1\pm D|V|)^{-4}$, with the plus (minus) sign for the retrograde (prograde) component. 
In other words, the prograde wave is more rapidly damped than the retrograde one. This statement is true even in the purely hydrodynamical case and it's notably this asymmetry that permits the development of the SLO despite a null total flux of angular momentum (i.e., the sum of both the prograde and retrograde components) injected at the top boundary. Using the mean flow equation in Eq.~(\ref{eqBrunoNS_okadim}) and taking explicitly the derivative of $\ds Rey$, we can estimate from heuristic arguments the timescale on which IGW can modify the mean flow. For the prograde component acting efficiently on the mean flow close to the base of the convective interface, it is given by $\ds \tau_P \sim V_0(1-D V_0)^4L/F$ (in code units), whereas for the retrograde component acting efficiently in deeper layers, we have $\ds \tau_R \sim V_0(1+DV_0)^4L/F$, with $V_0$ the typical velocity amplitude of the mean flow. The time needed for the retrograde component to act on the flow and efficiently create the shear is thus the largest one: it thus mostly determines the oscillation timescale. In the MHD case, we have seen that $V_0$ (and thus the value of $Ro$) is decreased. The retrograde component is thus expected to act more rapidly, noting in addition the large sensitivity of the wave damping to $\ds V$ to the power four. This simple analysis certainly explains the increase of frequency in the MHD case that is displayed in figure \ref{fig_bifu_f_Ro}($\ds a$).}
 
\section{Conclusion and discussion}
\label{section_discussion}


In the spirit of mean-field dynamos such as classical $\ds \alpha-\Omega$ loops, the toy model reported here describes how parameterised equations can account for the combined effects of the shear-layer oscillation induced by internal gravity waves and the Tayler-Spruit dynamo. Measuring mean-field coefficients on 3D DNS by the SVD technique, we have shown that the magnetic field can strongly affect the properties of the classical (hydrodynamical) shear-layer oscillation. The period of the flow or the successive onsets of phase-locking events leading to a chaotic state are indeed modified once the dynamo sets in, the latter strongly changing the transport of angular momentum. Being one dimensional, the equations reported here can be implemented at little computational cost and provide the community with rich behaviours concerning the dynamical regimes achievable and have possible applications concerning the evolution of stars. \tck{One could indeed modify the angular momentum conservation equation considered in stellar evolution models \citep[e.g.,][]{Charbonnel2005,Marques2013} to describe the action of the Maxwell stress, besides taking into account the evolution of the magnetic field through the induction equation. The wave generation at the interface between the radiative and convective zones could be done similarly as what was proposed in previous studies \citep[e.g.,][]{Kumar99,Lecoanet2013,PinconBG16}.}

\tck{We nevertheless reiterate that our results are preliminary \tck{and highly idealised, both concerning the dynamo implementation and the possibility of a SLO in a stellar context. We have however mentionned the arguments that led us to derive such a simple dynamo model (see section \ref{section_discussion_TS})}. A future confirmation through thorough DNS models taking into account both the action of IGW and a dynamo will nevertheless be necessary.}

\subsection{Limitations of our approach}

\tck{We now discuss the assumptions we took in order to build our reduced model. At the core of the current model lies the choice of forcing frequency and amplitude, namely the choice of $\omega/N$ and $F$. It is interesting to compare our value of $\omega/N$ with those reported in the literature for solar-like stars by \citet{Rogers2006b}, \citet{Alvan2014} or \citet{Pinccon2016}. It is nevertheless worth pointing out that these studies, whether numerical or theoretical, deal with a broad range of frequencies rather than our monochromatic case. If \citet{Rogers2006b} report only small values with $\omega/N$ ranging from $10^{-3}$ to $10^{-2}$, \citet{Alvan2014} and \citet{Pinccon2016}’s spectra extend up to $\omega/N\sim 0.4$. Our choice of $\omega/N$ is therefore in their range of frequencies. The same can be said about recent simulations by \citet{Daniel2025} where spectra peak close to the value of $\omega/N \sim 0.2$ we chose. Nevertheless, it is true that the estimated peak of $\omega/N$ is rather around $10^{-3}$ at the top of the RZ \citep[e.g.,][]{Press81}. As this value constrains the SLO and Tayler length scales, we chose to work with this rather large values in order to study the feasibility of the coupling between the SLO and TS dynamo in a context similar to previous numerical studies.}

\tck{On comparing forcing amplitudes, a useful tool highlighted by \citet{Daniel2025} in order to compare monochromatic and continuous forcing is the value of the parameter originally introduced by \citet{Plumb1978} which reads $\Lambda_1=\nu c \gamma/(\omega^4 F_J)= (Pr/F) \left(N/\omega\right)^3(\tau_{\kappa} \omega D)^2$. When sufficiently small, a mean flow can develop as this parameter compares viscous diffusion to wave forcing. In both the present 1D case and the top of the Sun's RZ (Table \ref{tab_param}), it is at most equal to $10^{-3}$ and in good agreement with what they reported in order to observe a mean flow develop.}

Besides \tck{the choice of forcing mechanism}, in the hope of extrapolating such simple models to actual astrophysical considerations, it is also important to \tck{discuss} some of the hypotheses that were used to derive the equations at hand, notably concerning the ordering of typical time scales. Based on the formalism developed by \citet{Spruit02}, the magnetic field that can develop must remain \textit{weak}, i.e. the Alfvén frequency must be small compared to the rotation of the star and the Brunt-Väisälä frequency. This weak field assumption must be completed by saying that the Alfvén frequency must also be small compared to the one of the waves \citep{KimM03} to work with a version of the SLO equations in which the retroaction of the field on the flow is simply taken into account in the Lorentz force and not in the Reynolds stress tensor \tck{\citep[see also][]{Rogers2010,Macgregor2011}}. 
\tck{If the Alfvén frequency is not small compared to the frequency of the waves, this could lead to a strong modification of the waves properties, such as their damping length and propagation \citep[e.g.][]{KimM03, Mathis2012}. Waves may then be become magneto-gravity modes \citep[e.g.][]{Fuller2015, Lecoanet2017, Rui2023, David2025} or vertically trapped by an azimuthal magnetic field. In this context, a natural extension of the present work would be to consider the modified expression of the Reynolds stress in the case where a strong azimuthal magnetic field can modify the waves properties, as is for instance done by \citet{Mathis2012}.}

Besides \tck{considerations on magnetic field}, the frequency of the waves must be smaller than $\ds N$ and in the mean time much bigger than the rotation of the star so as to neglect \tck{the influence of the Coriolis acceleration on the waves}.
This leads to a regime of validity which must verify:
\begin{equation}
    \omega_A \ll \Omega_o \ll \omega \ll N.
    \label{eq_validity}
\end{equation}
\tck{Translating these inequalities as a function of the dimensionless parameters of the toy model, we can write:
\begin{equation}
    \widetilde{B} \frac{Ek}{Pr} \frac{r_o}{d}\ll  1 \ll (\tau_\kappa \omega) \frac{Ek}{Pr} \left(\frac{r_o}{d}\right)^2 \;\mbox{and}\; 1\gg(\tau_\kappa \omega)^{2/3} L^{1/3} D \; ,
\end{equation}
where we have considered $\ds r_o$ for the radius in the expression of the Alfvén frequency.}
\tck{As an additional assumption, we also note that} when the conditions for instability are met (Eq.~\ref{eq:criteres TI}), the $\ds \alpha$ term used to regenerate poloidal field from toroidal one (Eq.~\ref{eqBrunoPOLO_okadim}) is implemented \textit{instantly}. The underlying assumption is that the relevant time for the instability to set in $\ds \tau_{TI}$ \citep{Pitts1985} is small compared to the typical period of the SLO, \tck{which is close on order of magnitude to the timescale associated with the transport by IGW, discussed at the end of section \ref{section_results}.}
This can be translated by the following inequality:
\begin{equation}
    \tau_{TI} = \frac{\Omega_o}{\omega_A^2} \ll \tau_{SLO}  \approx \frac{L\widetilde{V_0}}{F}\frac{d^2}{\kappa},
    \label{eq_crit_slo_omega}
\end{equation}

\noindent or as a function of the dimensionless parameters:

\begin{equation}
    \frac{Pr}{Ek}\ll \frac{L}{F}\widetilde{V_0}\widetilde{B}^2,
\end{equation}

\noindent which even though is verified in our simulations is not a strong one. Note that $\ds \Omega_o$ in Eq.~(\ref{eq_crit_slo_omega}) could be chosen differently as the growth rate $\ds \tau_{TI}$ was derived in a context where $\ds \Omega_o$ is a good order of magnitude of the dominant angular velocity for the radiative zone of the star. In our case, the differential rotation induced by the SLO can lead to rotational profile in the bulk of the flow more important than the value at $\ds r=r_o$. See for instance figure \ref{fig_bifu_f_Ro}(b) where the Rossby number exceeds unity.

These limitations add up to the ones mentioned in \citet{Renaud2020} for the hydrodynamical case. Considering the parameters used in the present paper, we can check \textit{a posteriori} that the inequalities Eqs.~(\ref{eq_validity}) and (\ref{eq_crit_slo_omega}) are in general verified in our simulations, except for the first inequality in Eq.~(\ref{eq_validity}) related to the weak field approximation of the Tayler-Spruit dynamo that can sometimes be slightly broken in our simulations when the magnetic field becomes large. 
In our simulations, the timescale ordering Eq.~(\ref{eq_validity}) is respected most of the time, but in a stellar environment such as the Sun's RZ, the rotation frequency cannot be neglected compared to the wave frequency (Table \ref{tab_param}), which means that future studies should focus on modifying the present work to account for inertial gravity waves. Checking the ordering in Eq.~(\ref{eq_crit_slo_omega}) is less obvious. In particular, realistic values of $\tau_\kappa \omega$, $D$ or $F$ are needed in future work to check to what extent the model remains valid in stellar physical conditions, and the toy model with its explicit limitations lays the foundation for a more thorough study. Some estimates of the parameters are given in \tck{Table \ref{tab_param}} for the Sun's RZ.

Moreover, \tck{it is worth recalling that the toy model, although already complex, nevertheless relies on simplistic physical assumptions.} Regarding the very simple $\ds \alpha$ function implemented, additional terms could be taken into account, such as $\ds \gamma$ pumping or even 2nd order mean-field effects related to the $\ds \beta$ coefficients. Meridional flows, Coriolis acceleration and the strong field limits could also have an important role as well as a continuous spectrum of waves \citep{Leard2020}. \tck{The same goes for the turbulence triggered by the dynamo studied in \citet{DanielPG23}}. There is therefore room for improvement, \tck{which will be subject to further work}. 
\tck{As to the necessary comparison with DNS, numerical setups considering both the convective and the radiative zones must be taken into account, as was for instance achieved by \cite{Couston2017}. \cite{Couston2018} later observed a reversing mean flow in the stably stratified layer whose generation was attributed to the turbulent motion in the convective layer. Such a reversing mean flow is similar to the SLO (or QBO). Adding magnetic field in this numerical setup would enable to explore the possibility of a TS dynamo to be driven by this differential rotation. Note however that this work was conducted in Cartesian geometry and in 2D. Several studies have considered the interaction between a convective and a radiative zone in spherical geometry \citep[e.g.,][]{Rogers2006,Alvan2014} in a hydrodynamical regime in the anelastic approximation, or with magnetic field \citep[e.g.,][]{Brun2022}. None of these simulations of full stars reported an oscillating mean flow, perhaps because the adopted diffusion coefficients are still orders of magnitude away from realistic values. The oscillation \tck{is nevertheless possible in polar geometry, as shown by \citet{Daniel2025}, where the results of \cite{Couston2018} are extended at low Prandtl number}.}

\subsection{Perspectives}

As presented in section \ref{section_results}, our toy model stands \tck{currently} for a first step in the theoretical efforts to describe how magnetic field can affect angular momentum transport and modify the flow. \tck{It is thus worth discussing the potential perspectives, which are mainly twofold}.

We can first question the possibility to test {\color{black} the formalism of} the present model with observations \tck{in the future}. As recalled in the introduction, asteroseismology stands for a unique technique to probe the internal rotation of stars. This method usually relies on the study of their resonant global modes, which may exhibit detectable high-amplitude peaks in the Fourier spectrum of observed light curves. For the Sun, we have detected thousands of oscillation modes, enabling us to unveil its mean rotation profile down to about $\ds 0.3~R_\odot$ (where $\ds R_\odot$ is the radius of the Sun) from the surface \citep[e.g.,][]{Thompson1996}. However, even with more than thirty years of seismic data, directly resolving spatially and temporally the SLO below the base of the solar convective envelope where internal waves are generated appears unfeasible given its expected typical period and small thickness \citep{TalonKZ02}. Information on a potential SLO in the Sun could be more likely obtained in an indirect way, that is, through for instance its impact on the mean core rotation or on the thermal structure  or mixing at the interface between the radiative and convective zones that can be probed by seismology. {\color{black} To do so, the present model would have to be implemented into a 1D stellar evolution code to predict its internal effect over the Sun's life.} This statement also applies for distant main sequence (burning hydrogen in their core) and red giant stars (evolved stars having exhausted hydrogen in their core) with masses comparable to the solar value, in which the mean core rotation rate can also be probed \citep{mosser12,Benomar2015,Ouazzani2019}.
For more massive stars, the configuration is reversed compared to the Sun, with a convective core and a radiative envelope between which the magnetised SLO {\color{black} might} develop. The oscillation spectrum of standing waves in these stars becomes more complex to interpret; nevertheless, the detection of a stochastic low-frequency variability may also represent an interesting probe of the the coupling between dynamo and SLO if they turn out to be related \citep[e.g.,][]{Bowmman2019}. Indeed, the origin of such a spectral signature is still debated. Progressive IGW generated by core convective motions and propagating towards the surface have been previously proposed as a potential explanation for this variability \tck{\citep{Aerts2015}}. \cite{Lecoanet2019} disclaimed this idea arguing that radiative damping in this frequency regime is too large to allow these waves for reaching the surface layers. Keeping in mind that the comparison between the observed spectra and their excitation models may be subject to tedious discussions (e.g., in this paper, the discrepancy in the wave amplitude is very sensitive to the model parameters), it is reasonable to mention that the presence of a well-developed SLO close to the core boundary in such stars, which was not included in \cite{Lecoanet2019}, could also modify the picture. Indeed, the presence of a strong shear mean flow close to the core is expected to reduce the damping of outwards internal waves and maybe enable them to reach the surface, potentially providing an explanation for this low-frequency wave activity. In such a case, the observations of the signature could give in turn information on the SLO.
\tck{For instance, if one were to be able to measure the amplitude of these waves or even some bursts in their activity with a typical periodicity, one might be able to relate them to the nature of the magnetic field and of the mean flow through simple diagnostics provided by such a toy model (e.g., by comparison with the outwards wave Reynolds flux or the dominant frequency reported in figure \ref{fig_bifu_f_Ro}($\ds a$)).} This being said, it might be premature at this stage to risk ourselves as to give some quantified predictions concerning applications for real stars, due notably to the simplicity of our toy model and the efforts still needed to extrapolate towards realistic regimes, as discussed above. \tck{Although currently speculative, these possibilities are nevertheless worth to be addressed in future works.}

Secondly, regarding a purely nonlinear approach, the variety of dynamical regimes notably described in figure \ref{fig_regime} raises some exciting questions for the future and will require further studies. An important aspect would be to be able to derive the same kind of weakly nonlinear expansion achieved by \citet{semin2018} to extend it to our setup. As the invariance $\ds V \rightarrow -V$ of the equations is broken by adding magnetic field, it raises the question of how does the Hopf bifurcation behave once the magnetic field is taken into account. Furthermore, if the threshold for transitioning to chaos is indeed modified, will the underlying scenario follow the one observed by \citet{Renaud2019} or, as is more likely, how will it be affected by the magnetic field?

\tck{Finally, the work considered here adds up to a long list of low-dimensional approaches in stellar MHD fluid dynamics as the one introduced in \citet{DanielPG23}; see for instance \citet{Knobloch1998} for a "0D" order model describing the coupling between the velocity and the field, or \citet{Tobias1998} for a 1D one with two magnetic modes. Our toy-model follows the tradition of such models by showing some generality. Even if as discussed in section \ref{section_results_MHD} it cannot be directly described by the one of \citet{DanielPG23}, we showed that the Tayler-Spruit dynamo originally obtained with an imposed large scale shear of a Couette problem can also be triggered by the flow generated by the SLO, i.e. the source of differential rotation needed can be quite general. Beyond the Tayler-Spruit dynamo, one could very well adopt the same method as the one presented here to parameterise different dynamo effects to investigate their long time impact on the evolution of a star.}

\section*{Acknowledgements}
This work was granted access to the HPC resources of MesoPSL funded by the Region Ile-de-France and the project Equip@Meso (reference ANR-10-EQPX-29-01) of the programme Investissements d'Avenir supervised by the Agence Nationale pour la Recherche. FD wishes to thank François Petrelis for fruitful discussions. LP acknowledges financial support from ``Programme National de Physique Stellaire'' (PNPS) of CNRS/INSU, France. CG acknowledges financial support from the French program `JCJC' managed by Agence Nationale de la Recherche (Grant ANR-19-CE30-0025-01).

\vspace{12pt}

\bibliographystyle{gGAF}
\bibliography{ref}
\vspace{12pt}

\appendices
\section{Singular Value Decomposition (SVD)}

\label{SVD}

\begin{figure}[t]
\begin{center}
\begin{minipage}{120mm}
{\resizebox*{12cm}{!}{\includegraphics{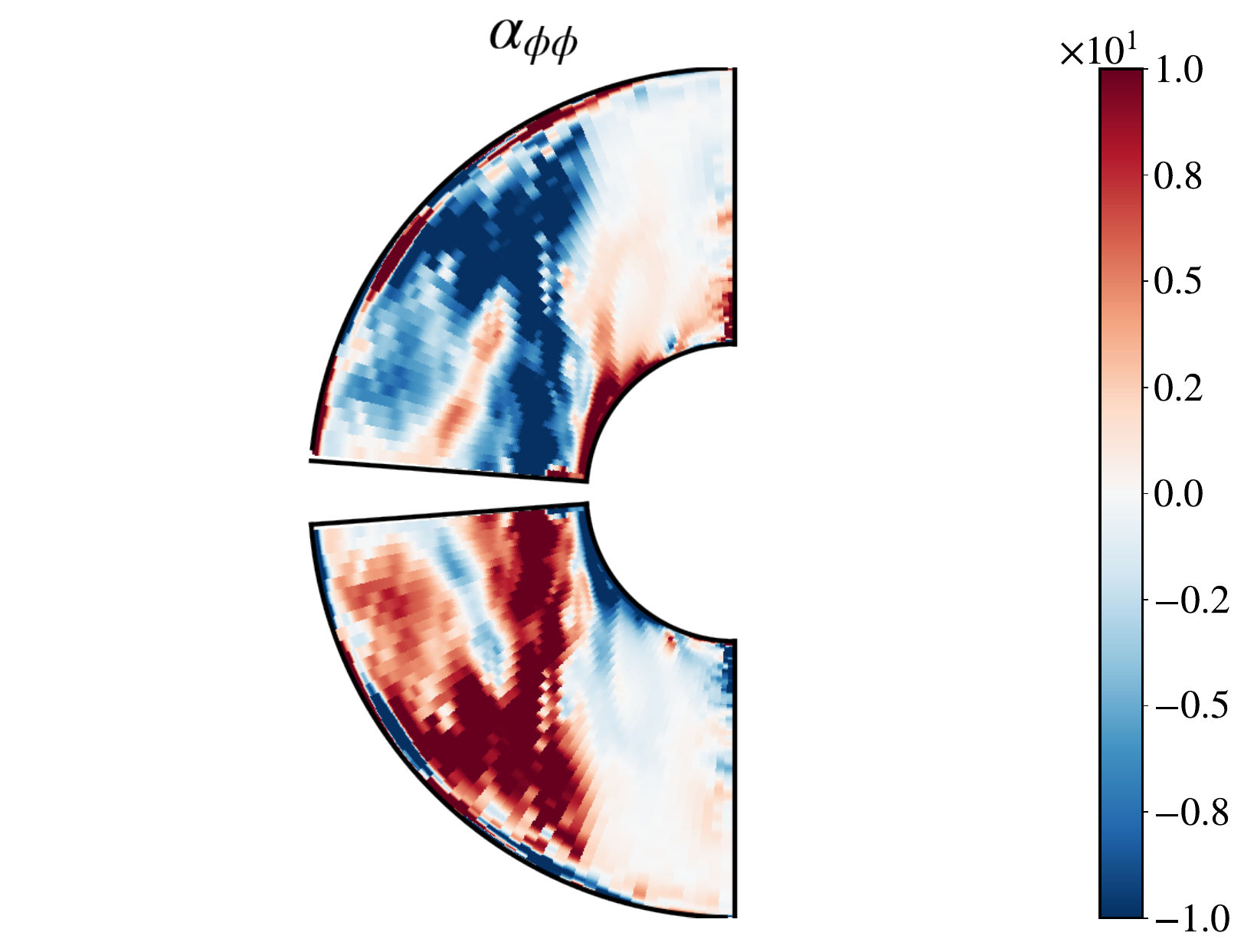}}}%
\caption{$\ds \alpha_{\phi\phi}$ obtained with the SVD approach explained here for a run reported in \citet{schrinner12} (model 4, see their figure 9). The SVD technique compares quite well to the test-field method as we retrieve the large scale component of $\ds \alpha_{\phi\phi}$ and its antisymmetric nature with respect to the equatorial plane. Close to the latter, the method did not converge probably due to a too little number of data files which the SVD relies on.}%
\label{fig_alpha_schrinner}
\end{minipage}
\end{center}
\end{figure}

Based on a time series approach, \citet{Racine2011,Simard2016,Dhang2020} bypass the $\textit{a priori}$ lack of equations in Eq.~(\ref{mean field}) - 3 equations for 9 $\ds a_{ij}$ unknown - by constructing the following matrix system:

\begin{equation}
    \mathcal{Y}(r,\theta)=\mathcal{A}(r,\theta)\mathcal{X}(r,\theta) + \hat{\mathcal{N}}(r,\theta),
    \label{schema1}
\end{equation}
where $\displaystyle \mathcal{Y} \in \mathcal{M}_{N,3}$, $\displaystyle \mathcal{A} \in \mathcal{M}_{N,3}$, $\displaystyle \hat{\mathcal{N}} \in \mathcal{M}_{N,3}$ and $\displaystyle \mathcal{X} \in \mathcal{M}_{3,3}$ is the unknown. $\displaystyle N$ is the number of steps of the system they consider, once the latter has reached a stationary state, where they assume $\displaystyle \mathcal{X}$ is constant. More precisely, $\displaystyle \mathcal{Y}$, $\displaystyle \mathcal{A}$ and $\displaystyle \mathcal{X}$ read, dropping the $\overline{\color{white}X\color{black}}$:

\begin{align}
    \nonumber &&\mathcal{Y} = \left( \begin{matrix} \mathcal{E}_{r t_1} & \mathcal{E}_{\theta t_1} & \mathcal{E}_{\phi t_1} \\ .. & .. & .. \\ \mathcal{E}_{r t_N} & \mathcal{E}_{\theta t_N} & \mathcal{E}_{\phi t_N}  \end{matrix} \right), \quad \mathcal{A} = \left( \begin{matrix} B_{r t_1} & B_{\theta t_1} & B_{\phi t_1} \\ .. & .. & .. \\ B_{r t_N} & B_{\theta t_N} & B_{\phi t_N}  \end{matrix} \right), 
    &&\mathcal{X} = \left( \begin{matrix} a_{rr} & a_{r\theta} & a_{r\phi} \\ a_{\theta r} & a_{\theta\theta} & a_{\theta\phi} \\ a_{\phi r} & a_{\phi \theta} & a_{\phi \phi}  \end{matrix} \right). 
\end{align}
Each line $\displaystyle i$ of Eq.~(\ref{schema1}) represents the state of the system at time $\displaystyle t_i$. $\displaystyle \hat{\mathcal{N}}$ is the numerical noise related to the expansion in Eq.~(\ref{mean field}). Eq.~(\ref{schema1}) is now over-determined, and an estimate of the coefficients of $\displaystyle \mathcal{X}$ is found by minimising:

\begin{equation}
        \chi_{i}^{2}(r, \theta)=\frac{1}{N} \sum_{n=1}^{N}\left[\frac{\left(\mathbf{y}_{i}\left(r, \theta, t_{n}\right)-\mathcal{A} \mathbf{x}_{i}\left(r, \theta, t_{n}\right)\right)^{\top}}{\sigma_{i}}\right]^{2},
        \label{squared_a_min}
\end{equation}
where $\displaystyle \mathbf{y}_i$ and $\displaystyle \mathbf{x}_i$ represents the $\displaystyle i^{th}$ column of $\displaystyle \mathcal{Y}$ and $\displaystyle \mathcal{X}$ respectively. $\displaystyle \sigma_{i}$ is the variance associated with the $\displaystyle i^{th}$ column of the noise matrix $\hat{\mathcal{N}}$. To seek Eq.~(\ref{squared_a_min}), the unique Singular Value Decomposition (SVD) of $\displaystyle \mathcal{A}$ is used:
\begin{equation}
    \mathcal{A}=\mathbf{U}\mathbf{w}\mathbf{V}^T,
\end{equation}
where $\displaystyle \mathbf{U}\in \mathcal{M}_{N,3}$, $\displaystyle \mathbf{V} \in \mathcal{M}_{3,N}$ and $\displaystyle \mathbf{w} \in \mathcal{M}_{3,3}$. $\displaystyle \mathbf{U}$ and $\displaystyle \mathbf{V}$ are orthonormal, while $\displaystyle \mathbf{w}$ is diagonal. The solution (i.e best fit) of Eq.~(\ref{squared_a_min}) $\displaystyle \hat{\mathbf{x_i}}$ can then be expressed as:
\begin{equation}
    \hat{\mathbf{x_i}}=\mathbf{V}\mathbf{w}^{-1}\mathbf{U}^T\mathbf{y}_i.
\end{equation}
The variance associated to each element of $\displaystyle \hat{\mathbf{x}}_i$ reads:
\begin{equation}
    \operatorname{Var}\left(\left[\hat{\mathbf{x}}_{j}\right]_{i}\right)=\sum_{l}\left[\frac{\mathbf{V}_{i l}}{\mathbf{w}_{l l}}\right]^{2} \sigma_{j}^{2},
\end{equation}
where $\displaystyle \sigma_j$ is calculated from the SVD fit as:
\begin{equation}
    \sigma_j^2=\frac{1}{N}\left(\mathbf{y}_j-\mathcal{A}\hat{\mathbf{x}}_j\right)^T\left(\mathbf{y}_j-\mathcal{A}\hat{\mathbf{x}}_j\right).
\end{equation}
We compare this approach to the test-field method in figure \ref{fig_alpha_schrinner}.

\section{\tck{Azimuthally-averaged angular momentum equation}}
\label{3Dto1Dflorentin}

To move from the 3D system to the 1D one, we start from the azimuthal component of the Navier-Stokes equation Eq.~(\ref{3:eq:EQ_TS_ADIM_1}) with dimension, multiply it by $r\sin \theta$ and average over the azimuthal direction,  $\ds \overline{X}=\frac{1}{2\pi}\int_0^{2\pi}X d\phi$. We follow the same idea as \citet{Eckhardt2007} in cylindrical coordinates. The result reads:

\begin{align}
    \nonumber &\frac{\partial }{\partial t} \overline{v_\phi r\sin\theta} + \overline{v_r r\sin \theta \pdr v_\phi} + \overline{v_r v_\phi \sin\theta } \\
    \nonumber + &\overline{v_\theta\sin\theta\pdth v_\phi } + \overline{v_\theta v_\phi \cos\theta}= \overline{ -2\Omega_o r\sin\theta \left(\cos\theta v_\theta + \sin\theta v_r\right)} \\
    \nonumber + &\overline{ V_{Ar} r\sin \theta \pdr V_{A\phi} } + \overline{ V_{Ar} V_{A\phi} \sin\theta } + \overline{V_{A\theta}\sin\theta\pdth V_{A\phi} }  \\
    \nonumber  + &\overline{ V_{A\theta} V_{A\phi} \cos\theta } + \nu\left(\overline{ \frac{\sin \theta}{r}\pdr \left(r^2\pdr v_\phi \right)} + \overline{ \frac{1}{r} \pdth \left( \sin \theta \pdth v_\phi\right)}\right. \\
    - &\left.\overline{\frac{v_\phi}{r\sin\theta}}\right),
    \label{eq:annexe1}
\end{align}
where each derivative with respect to $\ds \phi$  has disappeared over the azimuthal average (for instance the pressure term). We noted $\ds V_{Aj} = B_j/\sqrt{\rho \mu}$ for $\ds j=r,\theta,\phi$. We will in the following make use of the continuity equation Eq.~(\ref{3:eq:EQ_TS_ADIM_2}) which we will explicitly write in spherical coordinates:
\begin{equation}
    \frac{1}{r^2}\pdr (r^2 v_r) + \frac{1}{r\sin\theta}\pdth(\sin\theta v_\theta) + \frac{1}{r\sin \theta}\frac{\partial}{\partial \phi}v_\phi = 0.
    \label{eq:div v annexe}
\end{equation}
Focusing first on the inertial term on the left hand side of Eq.~(\ref{eq:annexe1}), we're going to rewrite the two terms involving $v_\theta v_\phi$:

\begin{align}
    \nonumber &\overline{v_\theta\sin\theta\pdth v_\phi } + \overline{ v_\theta v_\phi \cos\theta } \\
    \nonumber =&\overline{\pdth \left(v_\phi v_\theta \sin\theta \right)} - \overline{v_\phi \pdth \left(v_\theta \sin\theta\right)} + \overline{ v_\theta v_\phi \cos\theta }, \\
    \nonumber = &\overline{\pdth \left(v_\phi v_\theta \sin\theta \right)} +\overline{v_\phi \left( \frac{\sin\theta}{r}\pdr (r^2v_r)+\frac{\partial}{\partial \phi}v_\phi \right)} + \overline{ v_\theta v_\phi \cos\theta }, \\
\end{align}
where the continuity equation Eq.~(\ref{eq:div v annexe}) has been used. The term $\ds v_\phi\frac{\partial }{\partial \phi}v_\phi = \frac{\partial}{\partial \phi}\frac{v_\phi^2}{2}$ will give 0 once averaged. Finally, the total inertial contribution reads:
\begin{align}
    \nonumber &\overline{ v_r r\sin \theta \pdr v_\phi } + \overline{ v_r v_\phi \sin\theta } + \overline{v_\phi \frac{\sin\theta}{r}\pdr (r^2v_r)}+\overline{\pdth \left(v_\phi v_\theta \sin\theta \right)}+\overline{ v_\theta v_\phi \cos\theta } \\
    &=\frac{1}{r^2}\pdr \left(\overline{ v_r v_\phi} \sin\theta r^3\right)+\pdth \left(\overline{v_\phi v_\theta} \sin\theta \right)+\overline{ v_\theta v_\phi} \cos\theta.
\end{align}
The contribution arising from the Lorentz force has the same nature and reads:
\begin{align}
    \nonumber  &\overline{ V_{Ar} r\sin \theta \pdr V_{A\phi} } + \overline{ V_{Ar} V_{A\phi} \sin\theta } + \overline{V_{A\theta}\sin\theta\pdth V_{A\phi} } +\overline{ V_{A\theta} V_{A\phi} \cos\theta } \\
    \nonumber  & = \frac{1}{r^2}\pdr \left( \overline{V_{Ar} V_{A\phi}} \sin\theta r^3\right) + \pdth \left(\overline{V_{A\phi} V_{A\theta}} \sin\theta \right)+\overline{ V_{A\theta} V_{A\phi}} \cos\theta.
\end{align}
The Coriolis term simply reads:

\begin{align}
    \nonumber &\overline{ -2\Omega_o r\sin\theta \left(\cos\theta v_\theta + \sin\theta v_r\right)} = -2\Omega_o r\sin\theta \left(\cos\theta \overline{v_\theta} + \sin\theta \overline{v_r}\right),
\end{align}
while the viscous terms becomes:

\begin{equation}
    \nu\left( \frac{\sin \theta}{r}\pdr \left(r^2\pdr \overline{v_\phi} \right) + \frac{1}{r} \pdth \left( \sin \theta \pdth \overline{v_\phi}\right) 
    - \frac{\overline{v_\phi}}{r\sin\theta}\right).
    \label{eq_annexe_viscous}
\end{equation}

Before collecting every term, we're going to make the following assumptions: \begin{itemize}
    \item We assume that the latitudinal fluxes coming from the Reynolds and Maxwell stresses $\ds v_\phi v_\theta$ and $\ds V_{A\phi} V_{A\theta}$ respectively can be neglected compared to the radial ones $\ds v_\phi v_r$ and $\ds V_{A\phi} V_{Ar}$. This can be justified by saying that due to stable stratification, one expects the flow or the field spatial structures to be much more elongated in the latitudinal than in the radial direction and hence the transport more important in the latter than in the former case. 
    \item As discussed in section \ref{method}, we neglect the average of the meridional circulation $\ds \overline{v_r}$ and $\ds \overline{v_\theta}$ with respect to the zonal flow $\ds \overline{v_\phi}$.
    \item We make use of the local box approximation $\ds d \ll r_o$ which here translates in $\ds r\partial /\partial r \gg 1$. This holds if we set our local box not too close from the pole as the last term in Eq.~(\ref{eq_annexe_viscous}) would diverge there. 
\end{itemize} 

These assumptions enable us to eventually put Eq.~(\ref{eq:annexe1}) in the following form (dividing by $\ds r \sin \theta$) if we only take into account radial variations:

\begin{equation}
     \pdt \overline{ v_\phi}= - \pdr \left( \overline{v_r v_\phi} -\overline{V_{Ar}V_{A\phi}} \right) 
    +\nu \frac{\partial^2}{\partial r^2} \overline{v_\phi},
    \label{eq_annexe_conservation}
\end{equation}

\noindent which is the equation solved for $\ds V := \overline{v_\phi}$ in our 1d toy model.

\end{document}